\documentclass[12pt,a4paper]{article}
\usepackage{amsmath,amssymb,hyperref,bm,bbm}
\usepackage[utf8]{inputenc}
\numberwithin{equation}{section}
\allowdisplaybreaks[3]

\begin{document}
%%% Title page %%%%%
\begin{titlepage}
		
\renewcommand{\thefootnote}{\fnsymbol{footnote}}
\begin{flushright}
\begin{tabular}{l}
YITP-20-77\\
%\today %This should be commented out.
\end{tabular}
\end{flushright}
		
\vfill
\begin{center}
			
% \vskip 2.5 truecm
			
\noindent{\large \textbf{Correlation functions of symmetric orbifold}}
			
\medskip
			
\noindent{\large \textbf{from AdS$_3$ string theory}}
			
\vspace{1.5cm}

\noindent{Yasuaki Hikida\footnote{E-mail: yhikida@yukawa.kyoto-u.ac.jp} and Tianshu Liu\footnote{E-mail: u4842148@gmail.com}}
\bigskip
			
\vskip .6 truecm
\centerline{\it Center for Gravitational Physics, Yukawa Institute for Theoretical Physics,}
\centerline{\it  Kyoto University, Kyoto 606-8502, Japan}
% \medskip
			
\end{center}
		
\vfill
\vskip 0.5 truecm
		
\begin{abstract}
			
The paper examines correspondence among correlation functions of symmetric orbifold and string theory on AdS$_3$ described by $sl(2)$ Wess-Zumino-Novikov-Witten (WZNW) model. We start by writing down $n$-point function of twist operators in the symmetric orbifold in terms of the data of effective Riemann surface.  It is then shown that the correlation function can be reproduced from the $sl(2)$ WZNW model. The computation is based on the claim that string worldsheet is given by the same Riemann surface and the reduction method from $sl(2)$ WZNW model to Liouville field theory. We first consider the genus zero surface and then generalize the analysis to the case of generic genus. The result should be an important ingredient for deriving AdS$_3$/CFT$_2$ correspondence with tensionless superstrings to all orders in string perturbation theory.

\end{abstract}
\vfill
\vskip 0.5 truecm
		
\setcounter{footnote}{0}
\renewcommand{\thefootnote}{\arabic{footnote}}
\end{titlepage}
	
\newpage
	
\tableofcontents
%%%%%%%%%%%%%%%%%%%%%%%%%%%%%%%%%%%%%%%%%%%%%%%%%%%%%%%%%%%%%%%%%%%%%%

\section{Introduction}

It is widely believed that there appears an enhanced higher spin symmetry at the tensionless limit of superstring theory and the symmetry enables us to treat its stringy effects \cite{Gross:1988ue}.
At the same time, the tensionless limit of superstrings can be argued to be dual to a weakly coupled conformal field theory (CFT).
Therefore, we can make a quantitative check of AdS/CFT correspondence  including the stringy effects by making use of the tensionless limit.
Recently, it was claimed that a tensionless limit of superstrings on AdS$_3 \times S^3 \times T^4$ with NSNS-flux is exactly dual to the symmetric orbifold $T^4/S_N$.
As a support, the match of spectrum was already shown in  \cite{Gaberdiel:2018rqv,Eberhardt:2018ouy}, see also \cite{Giribet:2018ada}.
In this paper, we examine the correspondence of correlation functions in the duality described above, see \cite{Dei:2019osr,Eberhardt:2019ywk,Dei:2019iym,Eberhardt:2020akk} for previous works.
We first obtain the correlation functions of symmetric orbifold by following the method developed in \cite{Lunin:2000yv,Lunin:2001pw}. 
These correlation functions are then compared with those of AdS$_3$ string theory computed by applying findings of the previous works and the reduction method from AdS$_3$ string theory to Liouville field theory developed in  \cite{Hikida:2007tq,Hikida:2008pe}.

We consider a symmetric orbifold $\mathcal{M}^N/S_N$, where $\mathcal{M}$ is a generic two dimensional CFT with central charge $c$ and $S_N$ represents the symmetric group for the permutations of $N$ CFTs.
The symmetric orbifold includes twist operator $O_{(w)}(x)$ with twist number $w$ as a fundamental object,
and the correlation function of twist operators of the form
\begin{align}
\langle O_{(w_1)} (x_1) \cdots O_{(w_n)} (x_n) \rangle \label{corrSO}
\end{align}
is examined.
According to \cite{Lunin:2000yv}, this correlation function can be computed from the partition function of a single CFT $\mathcal{M}$ on a Riemann surface of genus $g$, which is defined by the covering map
\begin{align}
x = \Gamma (z) \label{xtoz}
\end{align}
 with a complex coordinate $z$. For the correlation function \eqref{corrSO}, the function $\Gamma (z) $ satisfies
\begin{align}
\Gamma (z_\nu ) \sim  x_\nu +  a_\nu^{\Gamma} (z - z_\nu )^{w_\nu}  \label{xtoz2}
\end{align}
around $z \sim z_\nu$ $(\nu=1,2,\ldots,n)$.

Generically speaking, it is a difficult problem to write down the covering map \eqref{xtoz} explicitly.
However, it is possible to fix the form of $\partial \Gamma(z)$ and the correlation functions can be written in terms of  its data.
As an illustration, let us consider a genus zero surface. In this case, we have
\begin{align}
\partial \Gamma (z) = \frac{ C  \prod_{\nu=1}^n (z - z_\nu)^{w_\nu -1}}{\prod_{\ell = 1}^R (z - y_\ell)^2} \label{dGammag0}
\end{align}
with
\begin{align}
R = \frac{1}{2} \sum_{\nu=1}^n (w_\nu - 1) + 1 \, . \label{Rg0}
\end{align}
Applying the procedure of \cite{Lunin:2000yv},
 we determine the correlation function \eqref{corrSO} as a function of $y_\ell$ as well as $z_\nu$ and $w_\nu$.

We would like to reproduce the correlation function from string theory.
Superstrings on AdS$_3 \times S^3 \times T^4$ with NSNS-flux can be described by a Wess-Zumino-Novikov-Witten (WZNW) model, and the AdS$_3$ part can be analyzed by the $sl(2)$ WZNW model
\cite{Giveon:1998ns,deBoer:1998gyt,Kutasov:1999xu}.
In the $sl(2)$ WZNW model, there is an operator $V^{w}_h (x;z)$ which is dual to the twist operator  $O_{(w)}(x)$ \cite{Maldacena:2000hw,Maldacena:2000kv,Maldacena:2001km}.
We thus evaluate the correlation function 
\begin{align}
\langle V_{h_1}^{w_1} (x_1 ; z_1) \cdots   V_{h_n}^{w_n} (x_n ; z_n) \rangle  \label{corrAdS}
\end{align}
in the $sl(2)$ WZNW model.
Here, $z$ is the worldsheet coordinate and $h$ is related to the conformal weight of $O_{(w)}(x)$.
Moreover, $w$ is so-called spectral flow parameter corresponding to the twist number $w$ of $O_{(w)}(x)$.
The parameter $x$ is introduced such as to be identified with the coordinate of symmetric orbifold.

For the evaluation of \eqref{corrAdS},
we utilize the claim that it  is localized at the Riemann surface defined by the same covering map $x = \Gamma(z)$  \cite{Eberhardt:2019ywk,Eberhardt:2020akk}.
In the case of sphere topology, the correlation function was proposed to take the form
\begin{align}
\langle V_{h_1}^{w_1} (x_1 ; x_1) \cdots   V_{h_n}^{w_n} (x_n ; z_n) \rangle 
= \sum_\Gamma \prod_{\nu=1}^n (a_\nu^{\Gamma} )^{- h_\nu} \prod_\nu \delta ^{(2)}(x_\nu - \Gamma (z_\nu)) W_{\Gamma} (z_\nu) \, , \label{EGG}
\end{align}
where  the sum is taken over all possible $\Gamma$ satisfying \eqref{xtoz2}.
This was confirmed by showing that \eqref{EGG} satisfies Ward-Takahashi identities in quite non-trivial ways.
We further apply the method developed in \cite{Hikida:2007tq,Hikida:2008pe} to reduce the correlation functions of the $sl(2)$ WZNW model involving non-trivial spectral flow parameters to those of Liouville field theory. The method was used to re-derive (generalized) Ribault-Teschner relation \cite{Ribault:2005wp,Ribault:2005ms}. See \cite{Hikida:2007sz,Creutzig:2011qm,Creutzig:2015hla,Creutzig:2020ffn} for  extensions of the work.

We first consider the Riemann surface of sphere topology and then generalize the analysis for higher genus surface.
It is known that the correlation functions in the symmetric orbifold $\mathcal{M}^N/S_N$ behave as \cite{Lunin:2000yv,Pakman:2009zz,Eberhardt:2019ywk}
\begin{align}
\langle O_{(w_1)} (x_1) \cdots O_{(w_n)} (x_n) \rangle \sim N^{1 - g - \frac{n}{2}} \, ,
\end{align}
when $x = \Gamma(z)$ defines a Riemann surface of genus $g$.
As mentioned above, we identify the Riemann surface with the worldsheet of dual string theory 
and examine the correspondence of correlation functions associated with the same Riemann surface.
In this manner, we would like to reproduce the $1/N$ corrections of correlation functions in the symmetric orbifold from the string perturbation theory with string coupling $g_s \sim N^{-1/2}$.

In order to analyze the correlation functions for supporting the duality with the tensionless limit of superstrings on AdS$_3 \times S^3 \times T^4$,  we need to set the level of $sl(2)$ WZNW model as $k=3$.%
\footnote{This limit also leads to a tensionless limit of bosonic strings on AdS$_3$ \cite{Gaberdiel:2017oqg}.}
However, it is also useful to consider bosonic strings on AdS$_3 \times X$ without taking any limit.
Here $X$ is a $23$-dimensional target space and the AdS$_3$ part is described by the $sl(2)$ WZNW model with generic level $k$. 
According to \cite{Eberhardt:2019qcl}, the string theory is related to the symmetric orbifold $\mathcal{M}^N/S_N$, where $\mathcal{M}$ is given by the product of Liouville field theory and the CFT for $X$ as
\begin{align}
\mathcal{M} =  \left[ \text{Liouville with } c= 1 + \frac{6 (k - 3)^2}{k-2}  \right] \times X \, . \label{bdual}
\end{align}
The central charge of $\mathcal{M}$ is
\begin{align}
c =  1 + \frac{6 (k - 3)^2}{k-2}  + 26 - \frac{3 k}{k -2 } = 6 k \, . \label{cbdual}
\end{align}
Though out the paper, we keep the level $k$ generic and realize that our findings are consistent with this type of relation as well.

 The organization of this paper is as follows.
 In the next section, we first explain the method developed in \cite{Lunin:2000yv} for constructing correlation functions in $\mathcal{M}^N/S_N$. 
 The method is then applied to obtain the correlation function of twist operators in symmetric orbifold  in terms of $y_\ell$ defined in \eqref{dGammag0} for the case where $x = \Gamma(z)$ defines a genus zero surface.
Section \ref{sec:sl2corr} aims computation of the correlation functions in the $sl(2)$ WZNW model with a sphere worldsheet.
For this purpose, we introduce the first order formulation of the WZNW model and vertex operators with non-trivial spectral flow parameter.
Correlation functions involving spectrally flowed operators are computed by applying the reduction procedure in \cite{Hikida:2007tq,Hikida:2008pe}. These correlation functions are then related to those in \eqref{corrAdS} by applying the findings of \cite{Eberhardt:2019ywk}.
 In section \ref{sec:hge}, we extend the analysis by working with the covering map \eqref{xtoz} defining a higher genus Riemann surface.
The section starts with a brief study on the basics of functions on a higher genus Riemann surface.
With these tools, we write down the correlation function of twist operators in symmetric orbifold in term of the positions of poles $y_\ell$ of $\Gamma(z)$. 
 We finally evaluate a correlation function of the $sl(2)$ WZNW model with the reduction method of \cite{Hikida:2007tq,Hikida:2008pe}, which was already developed on a higher genus Riemann surface.
 Section \ref{sec:conclusion} is devoted to conclusion and discussions.
 In appendix \ref{sec:beta}, the correlation functions of ghost system involving non-trivial spectral flow parameters are computed.
 
 \section{Correlation functions in symmetric orbifold} 
 \label{sec:SymOrb}

The goal of this section is to obtain correlation function  of twist operators in the symmetric orbifold $\mathcal{M}^N/S_N$ in a form that is useful for the comparison to the $sl(2)$ WZNW model.
The twist operator is denoted as $O_{(w)} (x)$ and its dimension is given by
\begin{align}
\Delta = \frac{c}{24} \left( w - \frac{1}{w}\right) \, . \label{SOtwist}
\end{align}
The method  of the computation 
is based on that developed in  \cite{Lunin:2000yv} but with a slightly modified regularization. 
The symmetric orbifold is defined on a manifold of sphere topology, and the metric is assumed to take the form
\begin{align}
ds^2 = | \rho (x) |^2 dx d \bar x \, , 
\end{align}
where we set $\rho (x) = 1$ in most cases. 
Denoting $X$ and $S[X]$ as the collection of fields and the action of them, respectively. The partition function for a single CFT $\mathcal{M}$ can be written as
\begin{align}
 Z_\delta = \int \mathcal{D} X e^{- S[X]}
\end{align} 
in the path integral formulation.
Here we have introduced an IR cut off at $x= 1/\delta \to \infty$.
The correlation function \eqref{corrSO} is then written as
\begin{align}
\langle O_{(w_1)} (x_1) \cdots O_{(w_n)} (x_n) \rangle  =  \frac{1}{Z_\delta^N}  \int \left [ \prod_{i=1}^N \mathcal{D} X_i  e^{- S [X_i] } \right ] O_{(w_1)} (x_1) \cdots O_{(w_n)} (x_n) 
\end{align}
with the insertion of  twist operators $O_{(w_\nu)} (x_\nu)$.
Here $X_i$ represents the collection of fields in the $i$-th CFT  $\mathcal{M}_i$.
In the presence of twist operator $O_{(w)} ( x )$, CFTs are exchanged as
\begin{align}
\mathcal{M}_{1} \to \mathcal{M}_{2} \, , \quad \mathcal{M}_{2} \to \mathcal{M}_{3} \quad  \ldots  \quad \mathcal{M}_{w} \to \mathcal{M}_{1} \, ,
\end{align}
when we go around the point $x $.

It is known that the path integral with the insertion of twist operators can be mapped to the partition function of a single $\mathcal{M}$ but on a non-trivial Riemann surface $\Sigma$. 
Namely, the correlation function \eqref{corrSO} can be given by
\begin{align}
\langle O_{(w_1)} (x_1) \cdots O_{(w_n)} (x_n) \rangle  = \frac{1}{Z_\delta^R} \int \mathcal{D} X e^{- S_\Sigma [X] } \, , \label{pf}
\end{align}
where $S_\Sigma[X]$ represents the action for $\mathcal{M}$ on $\Sigma$. 
Note that only $R$ of $N$ CFTs are involved, and the partition functions of $(N - R)$ CFTs left are canceled out \cite{Lunin:2000yv}. 
The number $R$ here is defined by \eqref{Rg0}.
The Riemann surface $\Sigma$ with a complex coordinate $z$ is defined by the covering map $x = \Gamma(z)$ satisfying \eqref{xtoz2}.
Around $z \sim z_\nu$, we have
\begin{align}
 z - z_\nu \sim ( a_\nu^\Gamma )^{-1/w_\nu} (x - x_\nu)^{1/w_\nu} \, . \label{zmznu}
\end{align}
This implies that we come back to the original point on $\Sigma$ with $z$-coordinate
when going $w_\nu$ times around the point $x_\nu$ in the $x$-plane.

Now the problem is to compute the partition function \eqref{pf} of a single $\mathcal{M}$ on the Riemann surface defined by the covering map $x = \Gamma(z)$.  The metric for the Riemann surface is given by
 \begin{align}
ds^2 = |\rho (x)|^2 d x d\bar x = \left| \frac{dx}{dz}\right|^2 |\tilde \rho(z)|^2 d z d \bar z  \, ,
\label{metric}
\end{align}
where $\tilde \rho(z) = \rho (\Gamma(z))$. 
We set $\rho (x)= \tilde \rho (z) = 1$ in general as we did before.
The metric \eqref{metric} is of the form 
\begin{align}
ds^2 = e^\alpha d \hat s ^2 \, , 
\end{align}
and the effect of $\alpha$ is removed by Weyl symmetry. 
However,  it is known that extra contribution arises due to the Weyl anomaly. 
We denote the partition functions computed with the metrics $ds^2 $ and $d \hat s^2$ by $Z^{(s)}$ and $Z^{(\hat s)}$, respectively.  Then, they are known to be related as \cite{Friedan:1982is}
\begin{align}
 	Z^{(s)} = e^{S_L} Z^{(\hat s)} \, , \label{Weylano}
\end{align}
where $S_L$ is the Liouville action given by
 \begin{align}
S_L = \frac{c}{48 \pi} \int d^2 z \left[ \partial \alpha\bar \partial \alpha + \frac{1}{2} \sqrt{g } \mathcal{R} \alpha \right] \, .
\label{Laction}
\end{align}
We choose the flat metric as $d \hat s = dz d \bar z$ and set $Z^{(\hat s)}  = 1$ since the comparison is made up to an overall factor.
In summary, the correlation function \eqref{corrSO} is obtained from  the on-shell Liouville action \eqref{Laction} with the field configuration
\begin{align}
\alpha = \ln | \partial \Gamma (z) |^2 \, .
\end{align}
However, we still need to choose a regularization to remove divergences arising during the computation.

We now proceed to explain our prescription where the covering map $x = \Gamma(z)$ defines a genus zero surface.
The extension to a higher genus surface is postponed to subsection \ref{sec:SymOrbg} after further preparations for basic tools.
In the case of genus zero, the derivative of $\Gamma(z)$ can be put in the form of \eqref{dGammag0}.
This means that the function $\Gamma(z)$ is given by integrating \eqref{dGammag0} over $z$ and it is 
fixed from the positions of poles $y_\ell$ up to the overall factor $C$  and an integration constant.
In general, it is a difficult problem to find out $y_\ell$ explicitly for given sets $(x_\nu,z_\nu)$, see, e.g., \cite{Lunin:2000yv,Pakman:2009zz}  for simple examples.
However, for the comparison of correlation functions, we only need the expression of the correlator \eqref{corrSO} in terms of the parameters $y_\ell$.

As explained above, we set
\begin{align}
 \alpha  = \ln \left|\frac{C\prod_{\nu =1}^n (z - z_\nu)^{w_\nu - 1}}{\prod_{\ell = 1}^R (z - y_\ell)^2} \right| ^2 
 \label{Lphi}
\end{align}
and evaluate the on-shell action of \eqref{Laction}. 
The factor $C$ corresponds to the scale of $x$-coordinate and we will set $C=1$ in the following by rescaling $x$.
Setting the curvature $\mathcal{R}=0$ due to the flat metric and performing partial integration, the Liouville action can be rewritten as%
\footnote{In order to properly treat the Liouville field theory on the flat metric, we need to add a boundary term as in \cite{Zamolodchikov:1995aa}. }
\begin{align}
 	S_L = - \frac{c}{48 \pi} \int d^2 z \alpha \partial \bar \partial \alpha \label{Lactionp} \, .
 \end{align}
The action of $\partial \bar \partial$ to \eqref{Lphi} can be written as
\begin{align}
\partial \bar \partial \alpha = 2 \pi  \sum_{\nu=1}^n (w_\nu -1) \delta^{(2)} (z - z_\nu)
- 4 \pi \sum_{\ell = 1}^R  \delta^{(2)} (z - y_\ell) \label{dbda}
\end{align}
using the identity 
\begin{align}
 	\partial \bar \partial \ln |z  - z '|^2 = 2 \pi \delta^{(2)} (z - z ') \, . \label{delta}
\end{align}
The on-shell action \eqref{Lactionp} is therefore given by the sum of contributions localized at $z = z_\nu$ and $z = y_\ell$.
However, we encounter a divergence for $\ln |z - z'|^2$ at the limit of $z \to z'$, and we have to decide how to regularize the divergence.

There are several ways to regularize this type of divergence, and a naive way is just to cut a hole with radius $\epsilon$ to remove the singular point as in \cite{Lunin:2000yv}.
However, the regularization depends on how to measure the regulator $\epsilon$ rather than how to regularize.
Since the Riemann surface defined by the covering map \eqref{xtoz} is not the real space but an effective one,
we should measure the length in the original coordinate system with the metric $ds^2 = dx d\bar x$.
A computation friendly regularization of the divergence
is adopted as in \cite{Polyakov:1981rd}, see also \cite{Hikida:2007tq}. 
Suppose that we are living on the $z$-plane, then we may regularize the divergence as
\begin{align}
\left( \lim_{z \to z'} \ln |z - z'|^2 \right)_\text{finite} = - \ln |\tilde \rho (z ' )|^2 = 0 \, . \label{reg0}
\end{align}
However, we are actually working with the $x$-plane, so we have to measure the length in terms of the $x$-coordinate.
This means that the correct regularization is
 \begin{align}
 	\left( \lim_{x \to x'} \ln |x - x'|^2 \right)_\text{finite} = - \ln |\rho (x ' )|^2 = 0 \, . \label{reg1}
 \end{align}
Near $z = z_\nu$, we thus use
\begin{align}
\begin{aligned}
\left( \lim_{z \to z_\nu} \ln |z - z_\nu|^2 \right)_\text{finite} 
&=\left( \lim_{x \to x_\nu} \ln  |(a^\Gamma_\nu )^{-1/w_\nu} (x - x_\nu)^{1/w_\nu} |^{2}  \right)_\text{finite} \\
&=  \frac{1}{w_\nu} \ln |a^\Gamma_\nu \rho (x_\nu) |^2  =  \frac{1}{w_\nu} \ln |a^\Gamma_\nu |^2 \, ,
\end{aligned} \label{reg2}
\end{align}
which is clearly different from \eqref{reg0}.

With the regularization of \eqref{reg1}, we evaluate the contributions to the on-shell action \eqref{Lactionp} localized at $z = z_\nu$ and $z=y_\ell$. 
Near $z = z_\nu$, the covering map $x = \Gamma(z)$ behaves as \eqref{xtoz2}
 with
 \begin{align}
a_\nu^\Gamma = \frac{ {\tilde a}_\nu^\Gamma}{w_\nu} \, , \quad 
{\tilde a}_\nu^\Gamma = \frac{ \prod_{\mu \neq \nu} (z_\nu - z_\mu)^{w_\mu - 1}}{\prod_{\ell} (z_\nu - y_\ell)^2}  \, .
  \label{aGamma}
 \end{align}
 Thus the contribution to the on-shell action is computed as
 \begin{align}
 	\begin{aligned}
 		S_L (z = z_\nu) &= - \frac{c}{24} (w_\nu -1 ) \lim_{z \to z_\nu } \ln |a_\nu^\Gamma w_\nu (z - z_\nu )^{w_\nu -1}|^2   \\
 		& = - \frac{c}{24} (w_\nu -1 )  \lim_{x \to x_\nu } \ln |(a_\nu^\Gamma)^{\frac{1}{w_\nu}} w_\nu (x - x_\nu )^{\frac{w_\nu -1}{w_\nu}}|^2 \\ 
 		&=  - \frac{c}{24} (w_\nu -1 )  \ln |(a_\nu^\Gamma)^{\frac{1}{w_\nu}} w_\nu |^2  \, .
 	\end{aligned}\label{contznu}
 \end{align}
 Here we have applied the regularization \eqref{reg2}. 
 Near $z = y_\ell$,  the covering map $x = \Gamma(z)$ behaves as
 \begin{align}
 	\Gamma (z) \sim - \frac{ \xi_\ell }{z - y_\ell } \, , \quad 
 	\xi_\ell =   \frac{ \prod_{\nu} (y_\ell - z_\nu)^{w_\nu - 1}}{\prod_{\ell ' \neq \ell} (y_\ell - y_{\ell ' })^2}  \, , \label{nearyell}
 \end{align}
 which will be derived shortly.
 The contribution to the one-shell action is
 \begin{align}
 	\begin{aligned}
 		S_L (z = y_\ell ) &= \frac{c}{12}  \lim_{z \to y_\ell} \ln \left | \frac{\xi_\ell}{(z - y_\ell)^2}\right|^2   \\
 		& =  \frac{c}{12}  \lim_{x = 1/\delta \to \infty}   \ln  | \xi_\ell ^{-1} x^2 |^2   \\ 
 		&=  - \frac{c}{12} \lim_{\delta \to 0} \ln  | \xi_\ell  \delta^{2} |^2 \, .
 	\end{aligned}\label{contyell}
 \end{align}
 
 Let us now go back to the expression of  $ \Gamma(z)$ near $z = \xi_\ell$ in \eqref{nearyell}.
Given the form of $\partial\Gamma(z)$ in \eqref{dGammag0}, we can write
 \begin{align}
 	\Gamma (z) = \frac{f_1 (z)}{f_2 (z)} \, , \quad f_2 (z) = \prod_{\ell = 1}^R (z - y_{\ell  }) \, ,
 \end{align}
from which we observe that the function $f_1(z) $ satisfies
 \begin{align}
 	f_2(z)^2 \frac{d \Gamma (z)}{dz} = \frac{d f_1(z)}{dz} f_2 (z) - f_1 (z ) \frac{d f_2 (z)}{dz} =  \prod_{\nu = 1}^n (z - z_\nu)^{w_\nu - 1} \, .
 \end{align}
 Setting $z = y_\ell$, we find
 \begin{align}
 	- f_1 (y_\ell)  \frac{d f_2 (y_\ell)}{dz}  =  \prod_{\nu = 1}^n (y_\ell - z_\nu)^{w_\nu - 1} \, ,
 \end{align}
 which leads to \eqref{nearyell}.

 Combining \eqref{contznu} and \eqref{contyell}, we arrive at%
 \footnote{This simple form was already suggested from the analysis of four point functions, see \cite{Lunin:2000yv,Pakman:2009zz,Roumpedakis:2018tdb,Dei:2019iym}.}
\begin{align}
\langle O_{(w_1)} (x_1) \cdots O_{(w_n)} (x_n) \rangle 
= \lim_{\delta \to 0}\frac{\delta^{- \frac{c R}{3}}}{Z_\delta^{R}}\prod_{\nu = 1}^n  |w_\nu |^{- \frac{c}{12} \frac{ (w_\nu - 1)^2}{w_\nu}} |  \tilde a_\nu^\Gamma|^{ - \frac{c}{12} \frac{w_\nu -1}{w_\nu}} 
\prod_{\ell = 1}^R |  \xi_\ell|^{- \frac{c}{6}} \, .  \label{SymOrb00}
\end{align}
We can derive $Z_\delta = \delta^{-c/3}$ up to a constant factor from \eqref{Weylano} with \eqref{Laction} as in (2.20) of  \cite{Lunin:2000yv}.
Further redefining twist operators as
\begin{align}
 |w_\nu |^{- \frac{c}{12} \frac{ (w_\nu - 1)^2}{w_\nu}} O_{(w_\nu)} (x_\nu) \to  O_{(w_\nu)} (x_\nu) \, , \label{Oredef}
\end{align}
the expression of \eqref{SymOrb00} becomes
\begin{align}
\langle O_{(w_1)} (x_1) \cdots O_{(w_n)} (x_n) \rangle 
= \prod_{\nu = 1}^n|  \tilde a_\nu^\Gamma|^{ - \frac{c}{12} \frac{w_\nu -1}{w_\nu}} 
\prod_{\ell = 1}^R | \xi_\ell|^{- \frac{c}{6}} \, .  \label{SymOrb0}
\end{align}

\section{Correlation functions in $sl(2)$ WZNW model}
\label{sec:sl2corr}

This section is devoted to the computation of correlation function of the form \eqref{corrAdS} in the $sl(2)$ WZNW model, which is expected to reproduce the same form as in \eqref{SymOrb0}.
The first subsection provides a review of the $sl(2)$ WZNW model with the symmetry of $sl(2)$ current algebra
and spectral flow automorphism of the algebra.
In subsection \ref{sec:sfo}, we examine the action of spectral flow to the vertex operators and introduce the parameter $x$ corresponding to the coordinate of dual symmetric orbifold.
In subsection \ref{sec:corrsfo}, correlation functions with spectrally flowed operators are computed using the reduction method developed in \cite{Hikida:2007tq,Hikida:2008pe}.
In subsection \ref{sec:relation}, we relate the correlation functions in subsection \ref{sec:corrsfo} with the correlation function of the form \eqref{corrAdS}.

\subsection{$sl(2)$ WZNW model and spectral flow}

As in \cite{Hikida:2007tq,Hikida:2008pe}, it is convenient to use the action of $sl(2)$  WZNW model in the first order formulation as
\begin{align}
S [\phi,\beta ,\gamma]= \frac{1}{2 \pi} \int d ^2 z \left(\partial \phi \bar \partial \phi - \beta \bar \partial  \gamma - \bar \beta \partial \bar \gamma + \frac{Q_\phi}{4} \sqrt{g} \mathcal{R} \phi - b^2 \beta \bar \beta e^{2 b \phi} \right) \, . \label{action}
\end{align}
The worldsheet metric and the scalar curvature are given by
\begin{align}
d s^2 = |\rho (z)| ^2 dz d \bar z \, , \quad \sqrt{g(z)} \mathcal{R}(z) = - 4 \partial \bar \partial \ln |\rho (z) |^2 \, ,
\end{align}
where we set $\rho(z)=1$ except at $z \to \infty$.
The background charge of $\phi$ is
$Q_\phi = b$, where we set $b^{-1} = \sqrt{k-2}$.
The central charge of the model is computed as
\begin{align}
c = 1 + 6 Q_\phi^2 + 2 = \frac{3 k}{k - 2} \, .
\end{align}
The symmetry algebra of the model is given by the $sl(2)$ current algebra,
whose generators satisfy the operator product expansions (OPEs),
\begin{align}
\begin{aligned}
&J^3 (z) J^3 (0) \sim -  \frac{k/2}{z^2 } \, , \quad
J^3 (z) J^{\pm} (0) \sim \pm \frac{J^{\pm} (0)}{z} \, , \\
&J^+ (z) J^- (0) \sim \frac{k}{z^2} - \frac{2 J^3 (0)}{z} \, .
\end{aligned}
\end{align}
The generators can be written  as
\begin{align}
\begin{aligned}
&J^+ (z) = -  \beta (z) \, , \quad
J^3 (z) =  - (\beta \gamma ) (z) + b^{-1} \partial \phi (z) \, , \\
&J^- (z) =  -  (\beta (\gamma \gamma) ) (z) + 2 b^{-1} ( \gamma \partial  \phi) (z)  - k \partial \gamma (z)
\end{aligned} \label{sl2free}
\end{align}
in terms of $\beta,\gamma$ and $\partial \phi$.
There is another set of $sl(2)$ currents $\bar J^a (\bar z)$ in the anti-holomorphic sector, 
and they are expressed by $\bar \beta, \bar \gamma$ and $\bar \partial \phi$ analogous to \eqref{sl2free}.

The $sl(2)$ current algebra possesses the spectral flow automorphism
\begin{align}
\rho^S(J_n^3) = J^3_n - \frac{k}{2} S \delta_{n,0} \, , \quad \rho^S (J^\pm_n) = J^{\pm}_{n \pm S} \, , \label{spectralflow}
\end{align}
where the mode expansions of  the $sl(2)$ currents are given by
\begin{align}
J^3 (z) = \sum_{n \in \mathbb{Z}} \frac{J^3_n}{z^{n+1}} \, , \quad J^\pm (z) = \sum_{n \in \mathbb{Z}} \frac{J^\pm_n}{z^{n+1}} \, .
\end{align}
We define a state $|S \rangle$, which is obtained by acting $S$ units of spectral flow to the vacuum state $|0 \rangle$.
The state satisfies
\begin{align}
\rho^S (J^a_n) |S \rangle = 0 
\end{align}
for $n \geq 0$ and $a = 3 , \pm$.
The state $|S \rangle$ is decomposable as
\begin{align}
|S \rangle = |S\rangle_{(\beta , \gamma)} \otimes |S\rangle_\phi \, , 
\end{align}
where the two component states satisfy
\begin{align}
\beta_{n+S}  |S\rangle_{(\beta , \gamma)} = \gamma_{n-S}  |S\rangle_{(\beta , \gamma)} = 0 \, , \quad
| S \rangle _\phi = e^{- \frac{S}{b} \phi } |0 \rangle_\phi  \label{spvacua}
\end{align} 
for $n\geq 0$.

\subsection{Spectrally flowed operators}
\label{sec:sfo}

Let us now introduce vertex operators, in particular,  those with the action of spectral flow.
There are several bases used to express the vertex operators.
In order to include the action of spectral flow, the expression in the $m$-basis turns out to be useful.
The vertex operators without the action of spectral flow satisfy the OPEs
\begin{align}
\begin{aligned}
&J^\pm(z) V^j_{m, \bar m} (0) \sim \frac{m\pm j}{z} V^j_{m\pm 1, \bar m} (0) \, , \quad J^3 (z) V^j_{m, \bar m} (0) \sim \frac{m}{z} V^j_{m, \bar m} (0) 
\end{aligned}
\end{align}
in the $m$-basis. The conformal weight $h^{j}_m$ of the operator  $V^{j}_{m , \bar m} $ is given by
\begin{align}
h^{j}_m = - \frac{j (j-1)}{k-2} \, ,
\end{align}
which is proportional to the eigenvalue of the second Casimir operator $C_2 = - j (j-1)$.
Note that the conformal weight is invariant under the exchange $j \leftrightarrow 1-j$.
The vertex operators are taken of the form
\begin{align}
V^j_{m, \bar m} =   \gamma^{- j  - m} \bar \gamma^{- j  - \bar m}  e^{2 b j \phi} 
\end{align}
in terms of fields appearing in the action \eqref{action}.

In order to explain the effect of spectral flow to the vertex operators,
it is convenient to move to the coset model $sl(2)/u(1)$ by decomposing $sl(2) \sim sl(2)/u(1) \oplus u(1)$.
Extra $u(1)$ currents are introduced as
\begin{align}
H (z) = - \sqrt{\frac{k}{2}} \partial \chi (z) \, , \quad \bar H (\bar z) = - \sqrt{\frac{k}{2}} \bar \partial \bar  \chi (\bar z) \label{H}
\end{align}
 with
\begin{align}
 \chi (z) \chi (0) \sim  - \ln z \, , \quad \bar  \chi (\bar z) \bar \chi (0)  \sim  - \ln  \bar z  \, .
\end{align}
The charged currents can be decomposed as
\begin{align}
 J^{\pm} = \Psi_\pm e^{\pm \sqrt{\frac{2}{k}}\chi } \label{currentdec}
\end{align}
in terms of parafermionic fields $\Psi_\pm$.
In an analogous manner, the vertex operator $V^{j}_{m , \bar m}$ can be decomposed as 
\begin{align}
V^{j}_{m , \bar m} 
=  \Psi^j_{m , \bar m} e^{ \sqrt{\frac{2}{k}} ( m \chi + \bar m \bar \chi) } 
\label{spectralm0}
\end{align}
with a coset vertex operator $\Psi^j_{m , \bar m} $.

We denote the vertex operator obtained by acting $w$ units of spectral flow on $V^j_{m , \bar m}$  as $V^{j,w}_{m , \bar m} $. The eigenvalue of $J_0^3$ can be read off from \eqref{spectralflow} as $m + kw/2$.
The vertex operator $V^{j,w}_{m , \bar m}$ is defined using the coset vertex operator $ \Psi^j_{m , \bar m} $  as (see, e.g., \cite{Argurio:2000tb})
\begin{align}
V^{j,w}_{m , \bar m} 
  =  \Psi^j_{m , \bar m}e^{ \sqrt{\frac{2}{k}} ( (m + \frac{kw}{2}) \chi + (\bar m + \frac{kw}{2}) \bar \chi ) }  \, . \label{spectralm}
\end{align}
The conformal weight $h^{j,w}_m$ of the operator  $V^{j,w}_{m , \bar m} $ is given by
\begin{align}
h^{j,w}_m = -\frac{j (j-1)}{k-2} - w m - \frac{k}{4} w^2 \, .
\end{align}
With this definition of  $V^{j,w}_{m , \bar m} $ and the expression of currents in \eqref{currentdec}, one can derive the following OPEs
\begin{align}
\begin{aligned}
&J^\pm (z) V^{j,w}_{m , \bar m}  (0)= \frac{(m\pm j)V^{j,w}_{m\pm 1 , \bar m}  (0)}{z^{\pm w+1}} + \mathcal{O} (z^{\mp w}) \, , \\
&J^3 (z) V^{j,w}_{m , \bar m}  (0)=  \frac{(m + \frac{k w}{2})V^{j,w}_{m , \bar m}  (0)}{z} + \mathcal{O} (z^{0}) \, . \\
\end{aligned} \label{JVOPEm}
\end{align}

Let us illustrate the discussion above with an example. We denote the operator corresponding to the state $|S \rangle$, which is constructed by acting $S$ units of spectral flow on the vacuum state, by $v^{(S)} (\xi)$. As in \eqref{spectralm}, the operator can be constructed by
\begin{align}
	v^{(S)} (\xi)  =  \Psi^{0}_{0,0}  e^{S \sqrt{\frac{k}{2}} (\chi (\xi) + \bar \chi (\bar \xi))}  \, . \label{vs}
\end{align}
The conformal weight of this operator is $h^{(S)} = - k S^2/4$. Notice that $ \Psi^{0}_{0,0} $ is the identity operator in the coset model.
Recall that the aim is to compute the correlation functions of vertex operators with the action \eqref{action}  by performing the path integral over the fields $\beta,\gamma$ and $\phi$. To serve this purpose, we shall decompose the operator  $v^{(S)} (\xi )$ as
\begin{align}
v^{(S)} (\xi) = v^{(S)}_{(\beta,\gamma)}(\xi) \otimes v^{(S)}_\phi (\xi) \, . \label{vs2}
\end{align}
From \eqref{spvacua}, we observe that the insertion of $v^{(S)}_{(\beta,\gamma)}(\xi)$ puts restriction to the path integral domain of $\beta(z)$ (or $\gamma(z)$) and
\begin{align}
 v^{(S)}_\phi (\xi) = e^{- \frac{S}{b} \phi (\xi , \bar \xi)} \, . \label{vsphi}
\end{align}
Following \cite{Hikida:2008pe}, we require that $\beta(z)$ has a zero of order $S$ for $S>0$ and a pole of order $|S|$ for $S <0 $ at $z = \xi$.

A generic operator $V^{j,w}_{m , \bar m}$ may be generated by the OPE between $v^{(w)}$ and $V^{j}_{m , \bar m}$.
The expression of $v^{(w)}  $ in \eqref{vs2} with \eqref{vsphi} implies that the operator takes the form
\begin{align}
V^{j,w}_{m , \bar m} = \rho^w (\gamma^{-j-m}) \rho^w (\bar \gamma^{-j - \bar m}) e^{2 b (j - \frac{w}{2 b^2}) \phi} \, .
\end{align}
The correlation functions involving the factors $\rho^w (\gamma^{-j-m}) $ and $\rho^w (\bar \gamma^{-j - \bar m}) $ will be obtained by relating to those of simpler forms.

Recall that the vertex operators in the correlation function \eqref{corrAdS} are in the $x$-basis, which is convenient for the application to AdS/CFT correspondence, since $x$ is identified with the coordinate of dual CFT \cite{deBoer:1998gyt,Kutasov:1999xu}.
The $x$-dependence in a spectrally flowed operator is introduced as (see, e.g.,  \cite{Maldacena:2001km,Eberhardt:2019ywk})
\begin{align}
V^w_h (x;z) =  e^{x J^+_0} V^w_h (0;z) e^{ - x J^-_0}  \, . \label{xpara}
\end{align}
We set the operator located at $x=0$ to be
\begin{align}
V^w_h (0;z) \equiv V^{j,w}_{m,\bar m} (z) \, . \label{Vh0}
\end{align}
On the left-hand side, we introduced a parameter
$
h = m + \frac{k w}{2}
$
 and suppressed  $\bar h = \bar m + k w/2$ and $j$. 
Since we can exchange $(m,w)\leftrightarrow (-m,-w)$ using the automorphism $(J^3 , J^\pm) \leftrightarrow (- J^3 , - J^\mp)$, we restrict the spectral flow number to $w \geq 0$. 
The OPEs between the $sl(2)$ currents and $V^w_h (x;z) $  are obtained as 
\begin{align}
J^a (z) V^w_h (x ; 0) = e^{x J_0^+} [ J^{a(x)} (z) V_h^w (0;0) ] e^{- x J^+_0}  \label{JVOPE2}
\end{align}
with
\begin{align}
\begin{aligned}
&J^{+(x)} = J^+ (z) \, , \quad J^{3 (x)} (z) = J^3 (z) + x J^+ (z) \, , \\
&J^{- (x)} = J^- (z) + 2 x J^3 (z) + x^2 J^+ (z) \, ,
\end{aligned} \label{Jax}
\end{align}
where the $x$-dependence is read off from \eqref{xpara}. The OPEs involving $V_h^w (0;0)$ are
\begin{align}
\begin{aligned}
&J^\pm (z) V^w_h (0;0) = \frac{(m\pm j)V_{h \pm 1}^w  (0;0 )}{z^{\pm w+1}} + \mathcal{O} (z^{\mp w}) \, , \\
&J^3 (z)  V^w_h (0;0) =  \frac{h V^w_h (0;0)}{z} + \mathcal{O} (z^{0}) \, , \\
\end{aligned} \label{JVOPE}
\end{align}
which are the same as \eqref{JVOPEm} due to \eqref{Vh0}.

Since the $sl(2)$ generators $J^3_0,J^\pm_0$ can be identified with a subset of space-time Virasoro generators $\mathcal{L}_0, \mathcal{L}_\mp$, we can regard the quantum number $h$ as the conformal dimension of dual operator. We may construct an worldsheet operator
\begin{align}
\mathcal{V}_h^w (x;z) = V_h^w (x;z) U_\Delta (z)  \label{UDelta}
\end{align}
by multiplying an operator $U_\Delta (z)$ with conformal dimension $\Delta$. The total conformal dimension must be one, and this yields the relation
\begin{align}
- \frac{j (j-1)}{k-2} - w m - \frac{k}{4} w^2 + \Delta = 1 \, . \label{onshell} 
\end{align}
Let us set $j=1/(2 b^2)$ (or $j = 1 - 1/(2b^2)$) and $U_\Delta$ to be the identity operator with $\Delta = 0$.
Then, the relation \eqref{onshell} leads to
\begin{align}
h = m + \frac{k w}{2} = \frac{k}{4} \left(w - \frac{1}{w}\right) \, . \label{htwist}
\end{align}
This reproduces \eqref{SOtwist} if we set $c = 6 k$ as in \eqref{cbdual}.

\subsection{Correlation functions with spectrally flowed operators} 
 \label{sec:corrsfo}

We would like to compute the correlation function with the vertex operators defined by \eqref{xpara} and \eqref{Vh0} as stated in \eqref{corrAdS}.
We are interested in the case where $h_\nu = m_\nu + k w_\nu /2 $ satisfies \eqref{htwist} and $j_\nu=1/(2b^2)$ (or $j_\nu =1 - 1/(2 b^2)$).
However, we temporarily set $m_\nu,j_\nu$ arbitrary except for the condition
\begin{align}
\sum_{\nu =1}^n j_\nu = \frac{n-2}{2b^2}  + 1 \label{conserve}
\end{align}
to be satisfied as was done in \cite{Eberhardt:2019ywk}. Note that \eqref{conserve} is satisfied for
\begin{align}
j_1 = 1 - \frac{1}{2b^2} \, , \quad j_\nu = \frac{1}{2 b^2} \label{jspecific}
\end{align}
with  $\nu = 2,3,\ldots , n$.

In this subsection, we evaluate a few correlation functions which take forms different from \eqref{corrAdS}.
As seen in the next subsection, these correlation functions reveal information on \eqref{corrAdS} while avoiding the difficulty of computing \eqref{corrAdS} directly.
We first examine
\begin{align}
\begin{aligned}
\tilde A_n &=\left \langle  v^{(-2)} (\xi) \prod_{\nu=1}^{n} V^{j_\nu  , 1 }_{m_\nu , \bar m_\nu} (z_\nu) 
\prod_{\ell = 1}^R  V^{\frac{1}{2b^2},1}_{\frac{k}{2} , \frac{k}{2}} (y_\ell) \right \rangle \\
&=\left \langle  v^{(-2)} (\xi) \prod_{\nu=1}^{\tilde n} V^{j_\nu  , 1 }_{m_\nu , \bar m_\nu} (z_\nu)\right \rangle
\end{aligned}
 \label{tildeA}
\end{align}
with $\tilde n = n + R$. In the second line of \eqref{tildeA},  we have set 
\begin{align}
V^{j_{n+\ell},1} _{m_{n+\ell}, \bar m_{n+\ell}} (z_{n+\ell}) =  V^{\frac{1}{2b^2},1}_{\frac{k}{2} , \frac{k}{2}} (y_\ell)
\end{align}
for $\ell = 1,2,\ldots , R$. 
For the moment, let us assume 
\begin{align}
\sum_\nu m_\nu = \sum_\nu \bar m_\nu = - \frac{k}{2}(2 R + n -2) \, . \label{conserve2}
\end{align}
Note that the correlation function is evaluated by the path integral with the action \eqref{action} in the first order formulation.

Taking account of the background charge $Q_\phi = b$, the momentum conservation for the $\phi$-direction is satisfied without the interaction term in \eqref{action}.
Here, we have used  \eqref{vsphi} and the condition \eqref{conserve}.
This, in particular, means that we can treat the $(\beta,\gamma)$-ghosts and the $\phi$-field separately
as
\begin{align}
\begin{aligned}
\tilde A_n 
&= \left \langle   v^{(-2)}_{(\beta,\gamma)}(\xi) \prod_{\nu=1}^{\tilde n} \rho^1 (\gamma ^{-j_\nu - m _\nu  } ) (z_\nu ) \rho^1 (\bar  \gamma ^{- j_\nu - \bar m_\nu } )  (\bar z_\nu )  \right \rangle_{(\beta, \gamma)} \\
& \quad \otimes 
\left \langle  e^{ \frac{2}{b} \phi (\xi , \bar \xi) } \prod_{\nu=1}^{ n}  e^{2 b ( j_\nu  - \frac{1}{2b^2} )\phi (z_\nu ,\bar z_\nu)} \right \rangle  _{\phi}  \, .
\end{aligned}
\end{align}
For  the $(\beta,\gamma)$-ghost part, the computation can be done by replacing the roles of $\beta$ and $\gamma$ as in appendix \ref{sec:beta}. The result is given by
\begin{align}
\begin{aligned}
&  \left \langle v^{(-2)}_{(\beta,\gamma)}(\xi) \prod_{\nu=1}^{\tilde n} \rho^1 (\gamma ^{-j_\nu - m _\nu  } ) (z_\nu ) \rho^1 (\bar  \gamma ^{- j_\nu - \bar m_\nu } )  (\bar z_\nu )  \right \rangle_{(\beta, \gamma)}\\
&\quad = \delta^{(2)} \left (\sum_{\nu =1}^{ n}   m_\nu +\frac{k }{2} (2 R + n -2) \right)   \prod_{\nu=1}^{\tilde n} (\xi - z_\nu)^{2 (j_\nu + m_\nu +1)} (\bar \xi - \bar  z_\nu)^{2 (j_\nu + \bar m_\nu +1)} 
\end{aligned} \label{onewinding}
\end{align}
up to some normalization factors.

Following \cite{Hikida:2007tq,Hikida:2008pe}, we shall perform some non-trivial manipulations for the $\phi$-part of the correlator, which in the path integral formulation is expressed as
\begin{align}
\left \langle  e^{\frac{2}{b} \phi   (\xi , \bar \xi) }  \prod_{\nu=1}^{ n} e^{2 b( j_\nu - \frac{1}{2b^2}) \phi (z_\nu , \bar z_\nu) } \right \rangle  _{\phi} 
= \int \mathcal{D} \phi e^{-S[\phi]}  e^{\frac{2}{b} \phi (\xi , \bar \xi)  }  \prod_{\nu=1}^{ n} e^{2 b(  j_\nu - \frac{1}{2 b^2} )\phi (z_\nu , \bar z_\nu) }  \, , 
\end{align}
where
\begin{align}
S[\phi] = \frac{1}{2 \pi} \int d ^2 z \left(\partial \phi \bar \partial \phi + \frac{Q_\phi}{4} \sqrt{g} \mathcal{R} \phi \right) \, .
\label{Lactionsl2} 
\end{align}
We then make a shift of $\phi$ as
\begin{align}
\phi + \frac{1}{2 b}  \left[ 2  \ln |\xi  - z |^2  + \ln |\rho (z)|^2 \right] \to \phi \, . \label{phishift}
\end{align}
This results extra terms contributed from the vertex operators and the kinetic terms of the correlator.
An analogous analysis as in \cite{Hikida:2007tq,Hikida:2008pe} leads to%
\footnote{Here we adopt the regularization \eqref{reg0} since we are living on the $z$-plane.}
\begin{align}
\left \langle  e^{\frac{2}{b} \phi  (\xi, \bar \xi)}    \prod_{\nu=1}^{ n} e^{2 b( j_\nu - \frac{1}{2b^2}) \phi  (z_\nu,\bar z_\nu)} \right \rangle  _{\phi} 
= \int \mathcal{D} \phi e^{-S[\phi]}\prod_{\nu=1}^{ n} e^{2 b ( j_\nu - \frac{1}{2b^2}) \phi (z_\nu , \bar z_\nu)}  |\xi - z_\nu|^{- 4 (j_\nu - \frac{1}{2b^2})}\, .
\end{align}	
The action is now \eqref{Lactionsl2} but with
\begin{align}
Q_\phi = b - b^{-1} \, , \label{Qnew}
\end{align}
which implies that the central charge of the theory is
\begin{align}
c = 1 + 6 Q_\phi ^2 = 1 + \frac{6 (k-3)^2}{k-2} \, . 
\end{align}
As we shall see below, one can identify the Liouville field theory as the one appearing in \eqref{bdual}.
Combining with \eqref{onewinding}, we find
\begin{align}
\label{tildeA2}
\begin{aligned}
\tilde A_n & = \int \mathcal{D} \phi e^{-S[\phi]}\prod_{\nu=1}^n e^{2 b ( j_\nu - \frac{1}{2b^2})  \phi (z_\nu , \bar z_\nu) } \prod_{\nu=1}^{\tilde n}(\xi - z_\nu)^{2 m_\nu + k} (\bar \xi - \bar z_\nu )^{2  \bar m_\nu + k}  \\
%& =  \delta^{(2)} (\sum_{\nu =1}^{ n}   m_\nu +\frac{k }{2} (2 R + n -2)) \prod_{\mu < \nu}  | z_{\mu \nu} | ^{ - 4 b^2 ( j_\mu  - \frac{1}{2b^2 }) ( j_\nu - \frac{1}{2b^2}) }  \, . 
\end{aligned}
\end{align}
under the conditions \eqref{conserve} and \eqref{conserve2}.

Let us now consider yet another correlation function
\begin{align}
A_n \equiv \left \langle \prod_{\nu=1}^n V^{j_\nu, w_\nu}_{m_\nu , \bar m_\nu} ( z_\nu)   \prod_{\ell = 1}^ {R}V^{\frac{1}{2b^2},-1}_{\frac{k}{2},\frac{k}{2}}(y_\ell) \right \rangle  \, . \label{assumption3}
\end{align}
As will be explained in the next subsection, this correlator is closely related to the one in \eqref{corrAdS}. 
Here we simply mention on the vertex operators  inserted at $z = y_\ell$.
It was argued in \cite{Eberhardt:2019ywk} that these vertex operators secretly exist in \eqref{corrAdS}.
Moreover, it is known that the sum of spectral flow parameters is bounded by the number $n$ of vertex operators inserted as 
$|\sum_ \nu w_\nu | \leq n -2 $ for a sphere correlation function.
For the case with $w_\nu > 0$ $(\nu = 1,\ldots,n)$, this bound is never satisfied.
However, the insertions of vertex operators at $z=y_\ell$ change both the number of vertex operators inserted and the sum of spectral flow parameters. This modifies the condition of the bound as
$
| \sum_{\nu}w_\nu - R | \leq R + n -2 \, .
$
This bound is saturated in this case due to \eqref{Rg0}. 
%Relatedly, for the case with the bound saturated, the conservation of $J_0^3$-charge becomes
%\begin{align}
%\sum_{\nu=1}^n m_\nu + \frac{k}{2} R = \sum_{\nu=1}^n \bar m_\nu + \frac{k}{2} R = - \frac{k}{2} (R + n -2) 
%\end{align} 
%as in \eqref{conserve2}. The condition may be written as
%\begin{align}
%\sum_{\nu =1}^n \left( m_\nu + \frac{k}{2} w_\nu \right) = \sum_{\nu=1}^n \left( \bar m_\nu + \frac{k}{2} \bar w_\nu \right) = 0  
%\end{align}
%since the extra vertex operators are set to have no $J_0^3$-charge.

Applying \eqref{spectralm}, we map the correlation functions with different spectral flow numbers but keeping the sum $\sum_\nu w_\nu$ invariant.  Via the correlation functions of the coset $sl(2)/u(1)$, we obtain the relation
\begin{align}
\left \langle \prod_{\nu=1}^n \Psi^{j_\nu}_{m _\nu , \bar m _\nu } (z_\nu) \prod_{\ell=1}^R \Psi^{\frac{1}{2b^2}}_{\frac{k}{2} , \frac{k}{2}} (y_\ell) \right \rangle = 
A_n |B_n| ^{-2}= \tilde A_n | \tilde B_n | ^{-2} \, . \label{BntBn}
\end{align}
Here, the $u(1)$ parts are given by
\begin{align}
B_n = \left \langle  \prod_{\nu=1}^n e^{ \sqrt{\frac{2}{k}} (m_\nu  + \frac{k w_\nu }{2} )\chi (z_\nu)} \right \rangle  \label{Bn}
\end{align}
and
\begin{align}
\tilde B_n =\left \langle  e^{ -  \sqrt{2k} \chi (\xi) }  \prod_{\nu=1}^n e^{ \sqrt{\frac{2}{k}} (m_\nu + \frac{k }{2} ) \chi (z_\nu)}  \prod_{\ell=1}^R e^{ \sqrt{2 k}  \chi (y_\ell) }  \right \rangle \, . \label{tBn}
\end{align}
Therefore, $A_n$ can be obtained as
\begin{align}
A_n = \tilde A_n |B_n  / \tilde  B_n|^{2} \, ,\label{An} 
\end{align}
where $\tilde A_n$ was computed in \eqref{tildeA2}.
Moreover, $B_n$ and $\tilde B_n$ can be evaluated by usual free field computations as
\begin{align}
 B_n / \tilde B_n &= \Theta \prod_{\mu < \nu} (z_{\mu \nu})^{ - m_\mu (w_\nu - 1) - m_\nu (w_\mu - 1) - \frac{k}{2} (w_\mu w_\nu - 1 )  } 
\prod_{\nu , \ell} (z_\nu - y_\ell)^{2 (m_\nu + \frac{k}{2}) } \prod_{\ell < \ell '} (y_{\ell \ell '})^{2 k} \nonumber \\
& = \Theta  \prod_{\nu} (\tilde a_\nu ^\Gamma )^{ -  m_\nu -  \frac{k }{4} ( w_\nu +1)} \prod_\ell \xi_\ell^{- \frac{k }{2} } 
\label{AdSresult}
\end{align}
with
\begin{align}
\Theta = \prod_{\nu=1}^n (\xi - z_\nu) ^{-2 (m_\nu + \frac{k}{2})}\prod_{\ell =1}^R (\xi - y_\ell)^{-2k} \, . \label{Theta}
 \end{align}

 In conclusion, we obtain
\begin{align}
\begin{aligned}
A_n & =  \prod_{\nu}  ( \tilde a_\nu ^\Gamma )^{ -  m_\nu -  \frac{k}{4} (w_\nu +1) }   ( \bar {\tilde a}_\nu ^\Gamma )^{ -  \bar m_\nu -  \frac{k}{4} (w_\nu +1) } \prod_\ell | \xi_\ell | ^{- k  } \\
& \quad \times \int \mathcal{D} \phi e^{-S[\phi]}\prod_{\nu=1}^n e^{2 b ( j_\nu - \frac{1}{2b^2})  \phi (z_\nu , \bar z_\nu) } \, , 
\end{aligned}
\label{Anresult}
\end{align}
where the action is \eqref{Lactionsl2} with $Q_\phi = b - b^{-1}$  as in \eqref{Qnew}.
Several comments on the result are in order.
Firstly, we can see that the $\xi$-dependent factor \eqref{Theta} cancels with that in \eqref{tildeA2}.
Secondly, the path integral over $\phi$ gives only a numerical constant when $j_\nu$ are given by \eqref{jspecific}.
Finally, compared to \eqref{SymOrb0}, there is still an extra factor
\begin{align}
\prod_{\nu} ( \tilde a_\nu ^\Gamma )^{ - m_\nu - \frac{k}{4} (w_\nu + \frac{1}{w_\nu})  } ( \bar {\tilde a}_\nu ^\Gamma )^{ - \bar m_\nu - \frac{k}{4} (w_\nu + \frac{1}{w_\nu})  } 
\end{align}
for $c = 6 k$ as in \eqref{cbdual}.
The factor becomes one for
\begin{align}
h_\nu = m_\nu + \frac{k w_\nu}{2} = \frac{k}{4} \left  (w_\nu - \frac{1}{w_\nu} \right) \, , \quad
\bar h_\nu = \bar m_\nu + \frac{k w_\nu}{2} = \frac{k}{4} \left(w_\nu - \frac{1}{w_\nu} \right)  \label{hwinding}
\end{align}
as desired.
However, this choice of $m_\nu$ and $\tilde m_\nu$ does not satisfy the condition \eqref{conserve2} though.
We will resolve  this issue in the next subsection.

\subsection{Relation to correlation functions in symmetric orbifold}
\label{sec:relation}

In the previous subsection, we evaluated the correlation function \eqref{assumption3}
 instead of \eqref{corrAdS}.
We now would like to show that the essential information of \eqref{corrAdS} can be obtained from \eqref{assumption3}.
In  \cite{Eberhardt:2019ywk}, it was claimed that the correlation function \eqref{corrAdS} should take the form of \eqref{EGG}, and this was confirmed by showing that the ansatz satisfies Ward-Takahashi identities.
%in quite non-trivial ways.
The ansatz \eqref{EGG} implies that the correlation function \eqref{corrAdS} takes a non-trivial value when parameters $(x_\nu,z_\nu)$ satisfy the relation $x_\nu = \Gamma(z_\nu)$. Moreover, we should take a sum over all possible covering maps $\Gamma(z)$. Here we would like to claim that the part proportional to the product of delta functions
$\prod_\nu \delta^{(2)} (x _\nu - \Gamma(z _\nu ))$ is obtained from  \eqref{assumption3}  once we fix the zero-mode of $\gamma$.

Before examining the relation of correlation functions, we would like to show that fixing the zero-mode of $\gamma$ effectively removes the condition \eqref{conserve2}.
The condition \eqref{conserve2} may be realized by the delta function in \eqref{onewinding}, which originates from the integration over parameter $u$ in \eqref{betacorr2}.
%Note that in \eqref{betacorr2}, the delta function is realized by the integration over parameter $u$. As we have discussed, \eqref{betacorr2} and \eqref{onewinding} are related by an overall factor. 
In the correlator of the form \eqref{onewinding}, $u$ is related to the zero-mode of $\gamma$.
%The parameter $u$ is, therefore, also related to the zero-mode $\gamma_0$ in \eqref{onewinding}.
Thus we need to take a constant $u$, which removes the delta function.
The value of $u$ can be absorbed by changing the normalization of vertex operators. 
The delta function from $(\beta,\gamma)$-ghost system is related to the conservation of $J^3_0$-charge, which is directly related to the conservation of the charge of  $H$-currents in \eqref{H}. Therefore, the condition \eqref{conserve2} is not necessarily satisfied if we fix the zero-mode of $\gamma$ and, in particular, we can safely set \eqref{hwinding} for all $\nu=1,2,\ldots,n$.

Our claim on the relation of correlation functions is supported by the fact, as we shall shortly show, that the two correlation functions  \eqref{assumption3}  and \eqref{corrAdS} share common properties.
We first examine some essential equations for correlation functions with the insertions of $\gamma$ and $\partial \phi$. 
In \cite{Eberhardt:2019ywk}, it was shown that the localization of the correlation function \eqref{corrAdS} at $x =\Gamma(z)$ can be seen in terms of the fields appearing the action \eqref{action} in the first order formulation.
For instance, it was claimed that the insertion of $\gamma(z)$ into the correlator \eqref{corrAdS} is replaced by the multiplication of the function $\Gamma(z)$:
\begin{align}
\left \langle \gamma (z) \prod_{\nu=1}^n V_{h_\nu}^{w_\nu} (x_\nu ; z_\nu) \right \rangle 
= \Gamma (z) \left \langle \prod_{\nu=1}^n V_{h_\nu}^{w_\nu} (x_\nu ; z_\nu) \right \rangle \, . \label{gammaWT}
\end{align}
In particular, $ \gamma (z)$ possesses a pole at $z = y_\ell$ even though there is no operator inserted there.
In a similar manner, the insertion of $\partial \phi (z)$ into the correlator leads to
\begin{align}
- b \left \langle \partial \phi (z) \prod_{\nu=1}^n V_{h_\nu}^{w_\nu} (x_\nu ; z_\nu) \right \rangle 
= \left( \sum_{\nu=1}^n \frac{b^2 j_\nu -  w_\nu / 2 }{z - z_\nu } + \sum_{\ell =1}^R \frac{1}{z - z_\ell}\right)\left \langle \prod_{\nu =1}^n V_{h_\nu }^{w_\nu } (x_\nu  ; z_\nu ) \right \rangle,  \label{dphiWT}
\end{align}
again with a pole located at $z = y_\ell$.
These equations were derived using the expressions of $sl(2)$ currents in terms of $\beta$, $\gamma$ and $\partial \phi$ and the fact that $sl(2)$ currents themselves should not have any pole at $z = y_\ell$.

The correlation function \eqref{assumption3} satisfies relations such as \eqref{gammaWT} and \eqref{dphiWT} as we will now show.
The second relation \eqref{dphiWT} tells us the information of momenta along $\phi$-direction for the vertex operators located at $z = z_\nu$ and $z=y_\ell$.
We can see that  the vertex operators in \eqref{assumption3} have the correct forms, and, in particular,
it fixes the quantum number $j=1/(2b^2)$ of $V^{j,w}_{m,\bar m}$ inserted at $z = y_\ell$.
Below we shall focus on the first relation \eqref{gammaWT}.

Around $z \sim z_\nu$, we can show that
\begin{align}
\label{gammaWT2}
 &\left \langle \gamma(z) \prod_{\nu=1}^n V^{j_\nu, w_\nu}_{m_\nu , \bar m_\nu} ( z_\nu)   \prod_{\ell = 1}^ {R}V^{\frac{1}{2b^2},-1}_{\frac{k}{2},\frac{k}{2}}(y_\ell) \right \rangle  \\
 &\sim c_\nu  \left \langle \prod_{\nu=1}^n V^{j_\nu, w_\nu}_{m_\nu , \bar m_\nu} ( z_\nu)   \prod_{\ell = 1}^ {R}V^{\frac{1}{2b^2},-1}_{\frac{k}{2},\frac{k}{2}}(y_\ell) \right \rangle + (z - z_\nu)^{w_\nu}  \left \langle \prod_{\nu=1}^n V^{j_\nu, w_\nu}_{m_\nu - 1 , \bar m_\nu - 1} ( z_\nu)   \prod_{\ell = 1}^ {R}V^{\frac{1}{2b^2},-1}_{\frac{k}{2},\frac{k}{2}}(y_\ell) \right \rangle \nonumber \, .
\end{align}
The first term with a constant $c_\nu$ arises from fixing the zero-mode of  $\gamma$ instead of integrating it out.  
The second term is obtained from the OPE between $\gamma$ and 
$ V^{j_\nu, w_\nu}_{m_\nu , \bar m_\nu} $, which can be deduced from, say, \eqref{sl2free} and \eqref{JVOPEm}.
Using \eqref{Anresult},%
\footnote{Precisely speaking, we  use the correlator \eqref{Anresult} with $\tilde a^\Gamma_\nu, \bar {\tilde a}^\Gamma_\nu$ replaced by $a^\Gamma_\nu, \bar a^\Gamma_\nu$. This can be realized by redefining vertex operators such as 
	$V^{j_\nu , w_\nu}_{m_\nu , \bar m_\nu} \to w_\nu^{m_\nu + \bar m_\nu + \frac{k}{2}(w_\nu +1)} V^{j_\nu , w_\nu}_{m_\nu , \bar m_\nu}$.} 
we find
\begin{align}
&\left \langle \gamma(z) \prod_{\nu=1}^n V^{j_\nu, w_\nu}_{m_\nu , \bar m_\nu} ( z_\nu)   \prod_{\ell = 1}^ {R}V^{\frac{1}{2b^2},-1}_{\frac{k}{2},\frac{k}{2}}(y_\ell) \right \rangle \nonumber  \\
&\quad \sim \left( c_\nu  + a_\nu^\Gamma (z - z_\nu)^{w_\nu}\right) \left \langle \prod_{\nu=1}^n V^{j_\nu, w_\nu}_{m_\nu , \bar m_\nu} ( z_\nu)   \prod_{\ell = 1}^ {R}V^{\frac{1}{2b^2},-1}_{\frac{k}{2},\frac{k}{2}}(y_\ell) \right \rangle, \label{gammaWT25}
\end{align}
which is consistent with (6.6) and (6.7) of  \cite{Eberhardt:2019ywk}.
Thus we reproduce the correct behavior of $\Gamma(z)$ as in \eqref{xtoz2} around $z \sim z_\nu$.
Recall  that the function $\Gamma(z)$ has a pole at $z = y_\ell$ as in \eqref{nearyell}.
In order to reproduce this behavior, we should include a vertex operator $ V^{j, w}_{m , \bar m} $ at $z = y_\ell$ with the OPE $\gamma(z) V^{j, w}_{m , \bar m} (y_\ell) \sim \mathcal{O} ((z - y_\ell)^{-1})$.
For this, we set $w=-1$ for the vertex operator at $z=y_\ell$ as in \eqref{assumption3}.%
\footnote{We set $m = \bar m = k/2$ for the vertex operator at $z = y_\ell$ such that the eigenvalues of $J_0^3$ and $\bar J_0^3$ vanish.}
Since the vertex operators in \eqref{assumption3} do not involve $\beta$ explicitly, we can integrate $\beta$ out, which yields $\bar \partial \gamma = 0$. This means that $\gamma$ should be replaced by a meromorphic function of $z$. 
Then the conditions of \eqref{gammaWT25} and the existence of first-order pole at $z = y_\ell$ lead to
\begin{align}
&\left \langle \gamma(z) \prod_{\nu=1}^n V^{j_\nu, w_\nu}_{m_\nu , \bar m_\nu} ( z_\nu)   \prod_{\ell = 1}^ {R}V^{\frac{1}{2b^2},-1}_{\frac{k}{2},\frac{k}{2}}(y_\ell) \right \rangle = \Gamma(z)  \left \langle \prod_{\nu=1}^n V^{j_\nu, w_\nu}_{m_\nu , \bar m_\nu} ( z_\nu)   \prod_{\ell = 1}^ {R}V^{\frac{1}{2b^2},-1}_{\frac{k}{2},\frac{k}{2}}(y_\ell) \right \rangle  \label{gammaWT3}
\end{align}
as in \eqref{gammaWT} up to an overall scaling and a constant shift.

So far, we have shown that relations \eqref{gammaWT} and \eqref{dphiWT} are satisfied by the correlation functions \eqref{assumption3}.
Using this fact, we shall clarify the relation between the correlation functions \eqref{assumption3} and \eqref{corrAdS}.
The vertex operators in \eqref{corrAdS} may be defined through the OPEs between $sl(2)$ currents as in \eqref{JVOPE2} with \eqref{Jax} and \eqref{JVOPE}.
We can see that the vertex operators $V^{j_\nu ,w_\nu}_{m_\nu , \bar m_\nu}$ inserted at $z = z_\nu$ satisfy the same OPEs inside the correlation function \eqref{assumption3}.%
\footnote{We also have to show that the $sl(2)$ currents $J^a(z)$ commute with  the vertex operator inserted at $z = y_\ell$ inside the correlation function \eqref{assumption3}.  However we have not completed the check for $J^-(z)$.} 
Namely, we can show that, around $z \sim z_\nu$,
\begin{align}
\left \langle J^a(z) \prod_{\nu=1}^n V^{j_\nu, w_\nu}_{m_\nu , \bar m_\nu} ( z_\nu)   \prod_{\ell = 1}^ {R}V^{\frac{1}{2b^2},-1}_{\frac{k}{2},\frac{k}{2}}(y_\ell) \right \rangle = \left \langle J^{a(x_\nu)}(z) \prod_{\nu=1}^n V^{j_\nu, w_\nu}_{m_\nu , \bar m_\nu} ( z_\nu)   \prod_{\ell = 1}^ {R}V^{\frac{1}{2b^2},-1}_{\frac{k}{2},\frac{k}{2}}(y_\ell) \right \rangle 
\end{align}
with $J^{a(x)}$ defined in \eqref{Jax}. 
The statement is trivial for $J^+(z)$.
For $J^3(z)$, we use its free field realization stated in \eqref{sl2free},
\begin{align}
J^3 (z) = - (\beta \gamma) (z) + b^{-1} \partial \phi (z) \, .
\end{align}
As in \eqref{gammaWT2} with $c_\nu = x_\nu$, $\gamma(z)$ in $J^3(z)$ gives two types of contribution inside the correlation function \eqref{assumption3}. Therefore, the insertion of $J^3 (z) $ can be effectively replaced by
\begin{align}
J^{3(x_\nu)} (z)= J^3 (z) + x_\nu J^+ (z) 
\end{align}
given in \eqref{Jax}. Similar arguments can be applied also for the insertion of $J^-(z)$.

In summary, the correlation function of twist operators in the symmetric orbifold $\mathcal{M}^N/S_N$ given in \eqref{corrSO} is shown to be reproduced from \eqref{corrAdS} in the $sl(2)$ WZNW model with conditions \eqref{jspecific} and \eqref{hwinding}.
Further correspondences among correlation functions can be deduced from this result.
Instead of \eqref{corrAdS}, we consider the correlation function of vertex operators defined in \eqref{UDelta} as
\begin{align}
\langle \mathcal{V}^{w_1}_{h_1} (x_1 , z_1) \cdots  \mathcal{V}^{w_n}_{h_n} (x_n , z_n) \rangle \label{gcorrAdS}
\end{align}
with generic $j_\nu$ but still subject to \eqref{conserve}.
The correlator exists in a bosonic string theory on AdS$_3 \times X$, and $U_{\Delta_\nu}$ is an operator in $X$ with $\Delta_\nu$ satisfying  \eqref{onshell}. The parameter $h_\nu$ is thus given by
\begin{align}
h_\nu = \frac{1}{w_\nu} \left( - \frac{\alpha_\nu (\alpha _\nu + k -3)}{k -2} + \Delta_\nu \right)  + \frac{k}{4} \left(w_\nu - \frac{1}{w_\nu}\right) \label{hnu}
\end{align}
with
\begin{align}
\alpha_\nu = j_\nu - \frac{1}{2b^2} \, .
\end{align}
As mentioned in the introduction, the string theory is supposed to be related to  the symmetric orbifold $\mathcal{M}^N/S_N$, where $\mathcal{M}$ is given by the product of Liouville field theory and $X$ as in \eqref{bdual}.
We can map \eqref{gcorrAdS} to the correlation function of the symmetric orbifold given by
\begin{align}
\langle \mathcal{O}^{\alpha_1}_{(w_1)} (x_1) \cdots \mathcal{O}^{\alpha_n}_{(w_n)} (x_n)  \rangle \label{gcorrSO}
\end{align}
with
\begin{align}
\mathcal{O}^{\alpha_\nu}_{(w_\nu)} (x_\nu) = O_{(w_\nu)} (x_\nu) e^{2 b \alpha_\nu  \phi (x_\nu , \bar x_\nu)} U_{\Delta_\nu} (x_\nu) \, .
\end{align}
Here the Liouville factor $e^{2 b \alpha_\nu \phi} $ arises due to \eqref{Anresult}.
This factor and the operator $U_{\Delta_\nu}$ in $X$ are now defined on the $x$-plane via $x = \Gamma(z)$. 
Using the covering map, we find that the zero-mode of Virasoro generator satisfies  
\begin{align}
\mathcal{L}_0 = \frac{1}{w_\nu} L_0 + \frac{c}{24} \left(w_\nu - \frac{1}{w_\nu} \right) \label{Virasororel}
\end{align} 
around $z \sim z_\nu$. The Virasoro generators defined in the $x$- and $z$-planes are denoted by $\mathcal{L}_m$ and $L_m$, respectively. When $c = 6k$, \eqref{Virasororel} yields a conformal weight for $\mathcal{O}^{\alpha_\nu}_{(w_\nu)} (x_\nu) $ which agrees with  \eqref{hnu}.
At the tensionless limit of $k=3$, all physical states are given by $j_\nu =1/2$, and the condition \eqref{conserve} is always satisfied as mentioned in \cite{Eberhardt:2019ywk}. However, for the correspondence with generic level $k$, the condition \eqref{conserve} is not always satisfied, and hence the correspondence of correlation functions has been shown only for limited cases.

\section{Higher genus extension}
\label{sec:hge}

In this section, we extend our previous analysis to higher genus Riemann surfaces.
Properties of Riemann surfaces with generic genus $g$ are reviewed in the next subsection.
In subsection \ref{sec:SymOrbg}, we compute the correlation function \eqref{corrSO} of the symmetric orbifold, where the covering map defines a higher genus Riemann surface.
In subsection \ref{sec:sl2g}, we examine the correlation function \eqref{assumption3} in the $sl(2)$ WZNW model with the higher genus worldsheet defined by the same covering map. They essentially compute \eqref{corrAdS} as in the genus zero case as the relation between \eqref{assumption3} and \eqref{corrAdS} can be established only from  local properties.

\subsection{Higher genus Riemann surface}
\label{sec:higherg}

This subsection is devoted to introducing useful facts about Riemann surfaces of generic genus $g$ which will serve our upcoming computation.
The conventions taken in this section follow from \cite{Verlinde:1986kw,Hikida:2007tq,Hikida:2008pe}, see also \cite{Fay:703829,Mumford:109230,AlvarezGaume:1986es}.
We denote a Riemann surface of genus $g$ by $\Sigma$ and introduce a complex structure.
There are $g$ numbers of holomorphic one-forms $\omega_l$ on $\Sigma$, and a canonical basis is chosen for homology cycles $\alpha_k , \beta_k$ such that
\begin{align}
\oint_{\alpha_k} \omega_l = \delta_{k,l} \, , \quad \oint_{\beta_k} \omega_l = \tau_{kl} \, . 
\end{align}
Here $\tau_{kl}$ is the period matrix of $\Sigma$.
In order to express functions on $\Sigma$, it is convenient to use the Abel map
\begin{align}
z_k = \int_{z_0}^z \omega_k \, ,  \label{Abel}
\end{align}
where $z_0$ is an arbitrary point in $\Sigma$. 
Adopting the Abel map, we introduce theta functions 
\begin{align}
\theta_\delta (z | \tau) = \sum_{n \in \mathbb{Z}^g} \exp\left\{  i \pi [(n+ \delta_1)^k \tau_{kl} (n+ \delta_1)^l + 2 ( n + \delta_1)^k (z+ \delta_2)_k ]\right\} \, .
\end{align}
Here $\delta_k  = (\delta_{1k} , \delta_{2k})$ with $ \delta_{1k},\delta_{2k} = 0,1/2$ represents the spin structure along the homology cycles $\alpha_k,\beta_k$.
The theta function has quasi-periodic property such that
\begin{align}
\theta_\delta (z + \tau n + m|\tau) = \exp [ - i \pi (n^k \tau_{kl} n^l + 2 n^k z_k)] 
\exp [ 2 \pi i (\delta^k_1 m_k - \delta_2^k n_k ) ] \theta_\delta (z|\tau) 
\end{align}
along the homology cycles. The Riemann vanishing theorem states that the theta function vanishes at a point $z$ if and only if there are $g-1$ points $p_i$ $(i=1,\ldots,g-1)$ such that $z$ can be written in a form
\begin{align}
z = \Delta - \sum_{i=1}^{g-1} p_i \, .
\end{align}
Here, $\Delta$ is a divisor class known as the Riemann class.

It is known that correlation functions of free boson theory can be expressed in terms of prime form,
\begin{align}
E (z,w) = \frac{\theta_\delta (\int^z_w \omega | \tau )}{h_\delta (z) h_\delta (w)} \, , 
\end{align}
with an odd spin structure $\delta$.
The auxiliary function $h_\delta(z)$ can be constructed through
\begin{align}
(h_\delta (z))^2 = \sum_k \partial_k \theta_\delta (0|\tau) \omega_k (z) \, .
\end{align}
The prime form has weight $(-1/2,0)$  both for $z$ and $w$ and has a zero as
\begin{align}
E(z ,w) \sim z -w \label{primezero}
\end{align}
at $z = w$.
It is periodic under the shift along $\alpha_k$ cycle but receives a non-trivial phase under the shift along $\beta_k$ cycle as 
\begin{align}
E (z + \tau_k , w) = - \exp \left( - i \pi \tau_{kk} - 2 \pi i \int^w_z \omega_k  \right) E(z,w) \, .
\end{align}
In the presence of background charge, we also need a $g/2$-form $\sigma(z)$ defined by 
\begin{align}
\ln | \sigma (z) |^2  =  \frac{1}{16 \pi } \int d^2 w \sqrt{g (w)} \mathcal{R} (w) \ln |E (z,w)|^2  \, ,  \label{sigmadef}
\end{align}
which satisfies
\begin{align}
\sigma (z + \tau_k , w) =  \exp \left( - \pi i (g-1) \tau_{kk}  + 2 \pi i \int^\Delta_{(g-1)z } \omega_k \right) \sigma (z  , w)  \, .
\end{align}

\subsection{Correlation functions in symmetric orbifold}
\label{sec:SymOrbg}

As reviewed in section \ref{sec:SymOrb}, the correlation function of twist operators \eqref{corrSO}  in the symmetric orbifold $\mathcal{M}^N/S_N$ can be obtained from a partition function of a single $\mathcal{M}$ but on a Riemann surface.
The Riemann surface is defined by the covering map $x = \Gamma(z)$ satisfying \eqref{xtoz2}.
In section \ref{sec:SymOrb}, we computed the partition function with a genus zero Riemann surface in terms of the positions of poles of $\partial \Gamma (z)$ given in \eqref{dGammag0}.
In this subsection, we examine the case with a Riemann surface of generic genus by extending the genus zero analysis.

We first express $\partial \Gamma(z)$ for the covering map $x = \Gamma(z)$ defining a higher genus Riemann surface $\Sigma$ as in \eqref{dGammag0}.
A meromorphic function on $\Sigma$ can be determined by its positions of zeros and poles.
We are looking for a function $\Gamma(z)$ satisfying \eqref{xtoz2}.
This condition fixes $\partial \Gamma(z)$ almost uniquely as 
\begin{align}
	\partial \Gamma (z) =  \frac{C \prod_{\nu=1}^n E(z , z_\nu)^{w_\nu -1} \sigma(z)^2 }{\prod_{\ell=1}^R E(z , y_\ell)^2 } \, . \label{dGammag}
\end{align}
Note that the prime form $E(z,w)$ has a zero at $z = w$ as in \eqref{primezero}.
The positions of poles are represented by $y_\ell$ and its number $R$ is given by the Riemann-Hurwitz formula as
\begin{align}
	R = \frac{1}{2} \sum_{\nu} (w_\nu -1) + 1 - g \, , \label{Rg}
\end{align}
which replaces the formula for the genus zero case in \eqref{Rg0}. The factor $\sigma (z)^2$ is inserted such that $\partial \Gamma(z)$ becomes a one-form. 
In order for $\partial \Gamma(z)$ to be a single-valued one-form on $\Sigma$, 
the positions of poles denoted by $y_\ell$ should satisfy
\begin{align}
\sum_{\nu=1}^n (w_\nu -1) \int_{z_0}^{z_\nu} \omega_l - 2 \sum_{\ell=1}^R \int_{z_0}^{y_\ell} \omega_l  - 2 \int_{(g-1)z_0}^\Delta \omega_l = 0 \, .
\end{align}
Note that the condition does not depend on $z_0$.
The factor $C$ corresponds to the scale factor of $x$, and we set $C=1$ in what follows as we did before. The function $\Gamma(z)$ is then obtained by integrating \eqref{dGammag} with respect to $z$.
%, but the integration constant will not appear in the final expression.

The on-shell action \eqref{Lactionp} can be evaluated analogously to the genus zero case in section \ref{sec:SymOrb}.
The Louville field in the higher genus case takes the form
\begin{align}
\alpha  = \ln \left| \frac{\prod_{\nu=1}^n E(z , z_\nu)^{w_\nu -1}  \sigma(z)^2  \tilde \rho(z)  }{\prod_{\ell=1}^R E(z , y_\ell)^2} \right|^2, \label{Lphig}
\end{align}
which reduces to \eqref{Lphi} for $g=0$. 
We mainly set $\rho(x) = \tilde  \rho(z) = 1$ but left $\tilde \rho(z)$ as it is when  it is necessary to keep track of the dependence of 
$ \partial \bar \partial \ln |\tilde \rho(z)|^2 $.
Acting $\partial \bar \partial $ to $\alpha$ in \eqref{Lphig} yields 
\begin{align}
\partial \bar \partial \alpha = 2 \pi  \sum_{\nu=1}^n (w_\nu -1) \delta^{(2)} (z - z_\nu)
- 4 \pi \sum_{\ell = 1}^R  \delta^{(2)} (z - y_\ell)  
+ \partial \bar \partial \ln |\tilde \rho (z) |^2  \, , \label{dbdag}
\end{align}
since $E(z,w)$ has only a zero at $z=w$ and behaves as \eqref{primezero} near the point.
Note that $\partial \bar \partial \ln |\sigma (z)|^2 = 0 $ since $\sigma (z)$ does not have any poles nor zeros.
Again, we adopt the regularization measured by the scale of $x$-coordinate as in \eqref{reg1}.

As in the genus zero case, there are contributions localized at $z = z_\nu$ and $z = y_\ell$ due to the delta functions in \eqref{dbdag}. The contribution to the action at $z = z_\nu$ is computed as
\begin{align}
	S_L (z = z_\nu) = - \frac{c}{24} (w_\nu -1) \ln | ( a^\Gamma_\nu )^{\frac{1}{w_\nu}} w_\nu  |^2 
\end{align}
just as in \eqref{contznu}.
The function $a^\Gamma_\nu$ is though modified as 
\begin{align}
	a^\Gamma_\nu = \frac{{\tilde a}^\Gamma_\nu \sigma (z_\nu)^2}{w_\nu} \, , \quad {\tilde a}^\Gamma_\nu = \frac{\prod_{\mu \neq \nu} E(z_\nu , z_\mu)^{w_\mu -1}   }{ \prod_{\ell} E(z_\nu ,y_\ell)^2} \, .
\end{align}
Similarly, the contribution at $z = y_\ell$ is
\begin{align}
	S_L (z = y_\ell) = - \frac{c}{12} \lim_{\delta \to 0} \ln  | \xi_\ell  \delta^{2} |^2\, , 
\end{align}
where we have used
\begin{align}
\Gamma (z) \sim - \frac{ \xi_\ell }{z - y_\ell } \, , \quad   	\xi_\ell = \tilde \xi _\ell  \sigma (y_\ell)^2 \, , \quad \tilde \xi_\ell = \frac{ \prod_{\nu} E(y_\ell , z_\nu)^{w_\nu -1} }{\prod_{\ell ' \neq \ell}E(y_\ell , y_\ell ')^2 } \, . \label{nearyellg}
\end{align}
This relation can be derived by arguments similar to the genus zero case.
The last term in \eqref{dbdag} and the identity $ \partial \bar \partial \ln |\tilde \rho(z)|^2  =  - \frac{1}{4} \sqrt{g(z)}\mathcal{R}(z)$ lead to
\begin{align}
\begin{aligned}
	 &	\frac{c}{192 \pi} \int d^2 z \sqrt{g(z)} \mathcal{R}  (z) \ln \left| \frac{\prod_{\nu=1}^n E(z , z_\nu)^{w_\nu -1} \sigma(z)^2 }{\prod_{\ell=1}^R E(z , y_\ell)^2  } \right|^2  \\
		&\quad =	 \frac{c}{12}\left[  \sum_{\nu=1}^n (w_\nu -1) \ln |\sigma (z_\nu)|^2 - 2 \sum_{\ell=1}^R \ln |\sigma (y_\ell)|^2 + \frac{3}{2} U_g \right]  \, ,
\end{aligned}
\end{align}
where $\sigma (z)$ was given in \eqref{sigmadef}, and the integral
\begin{align}
U_g = \frac{1}{192 \pi^2} \int d^2 z d^2 w \sqrt{g(z)} \mathcal{R} (z) \sqrt{g(w) } \mathcal{R} (w) \ln |E (z,w)|^2 
\end{align}
was used in the derivation.

Combining the above three types of contributions, we find
\begin{align}
	\begin{aligned}
		&\langle O_{(w_1)} (x_1) \cdots O_{(w_n)} (x_n) \rangle  \\
		&\quad =e^{\frac{c}{8} U_g} \prod_\nu |\tilde a_\nu^\Gamma |^{ - \frac{c}{12} \frac{w_\nu -1}{w_\nu}}  
		|\sigma (z_\nu )|^{\frac{c}{6} \frac{(w_\nu -1)^2}{w_\nu}}
		\prod_{\ell} |\tilde \xi_\ell |^{- \frac{c}{6}} | \sigma (y_\ell)|^{- \frac{2 c}{3}}  \, .
	\end{aligned}\label{SymOrb0g}
\end{align}
Note that the twist operators in this expression are redefined as in \eqref{Oredef}.
Moreover, similar to the genus zero case, the factor with $\delta$ in the contribution at $z=y_\ell$ is canceled by $Z_\delta^R = \delta^{- \frac{c R}{3}}$ in \eqref{pf}.
We will neglect the factor $e^{\frac{c}{8} U_g} $ in future discussions.

\subsection{Correlation functions in $sl(2)$ WZNW model}
\label{sec:sl2g}

The aim of this subsection is to reproduce the correlation function \eqref{corrSO} in symmetric orbifold from \eqref{corrAdS} in the $sl(2)$ WZNW model for higher genus Riemann surfaces by generalising the $g=0$ case discussed in section \ref{sec:sl2corr}.
Some of the arguments are based on local properties which can be directly applied to the higher genus case.
For instance, the correlation function \eqref{corrAdS} can be obtained from \eqref{assumption3} by removing the condition for the sum of $J_0^3$-charge. 
In this subsection, we first evaluate the higher genus analogue of \eqref{tildeA} by applying the reduction method generalized for the higher genus case in \cite{Hikida:2007tq,Hikida:2008pe}.
We then compute a correlator of the form \eqref{assumption3} using  the definition of spectrally flowed operators  in \eqref{spectralm}.

The correlation function of our interests takes the form as in \eqref{corrAdS}.
As discussed in \cite{Eberhardt:2019ywk,Eberhardt:2020akk}, we require that 
\begin{align}
\sum_{\nu =1}^n j_\nu = \frac{1}{2 b^2} (n-2 + 2 g) + 1 - g \, ,
\label{conserveg}
\end{align}
which reduces to \eqref{conserve} for $g=0$.
With this condition, the higher genus analogue of \eqref{tildeA}  is found to be
\begin{align}
	\tilde A_n =\left \langle  v^{(2g-2 )} (\xi) \prod_{\nu=1}^{\tilde n} V^{j_\nu  , 1 }_{m_\nu , \bar m_\nu} (z_\nu)\right \rangle \label{tildeAg}
\end{align}
with $\tilde n = n + R$ and
\begin{align}
	V^{j_{n+\ell},1} _{m_{n+\ell}, \bar m_{n+\ell}} (z_{n+\ell}) =  V^{1/(2b^2),1}_{\frac{k}{2} , \frac{k}{2}} (y_\ell)
\end{align}
for $\ell = 1,2,\ldots , R$. 
We further require the condition \eqref{conserve2} to hold for
  \eqref{tildeAg}.

Again, the correlator \eqref{tildeAg} is computed with the action of $sl(2)$ WZNW model in the first order formulation given in \eqref{action}.
As in the genus zero case, the momentum conservation along the $\phi$-direction is satisfied without using the interaction term in the action. In order to show this, we need to use \eqref{vsphi} and \eqref{conserveg}.
Thus the correlation function can be factorized as
\begin{align}
	\begin{aligned}
		\tilde A_n 
		&= \left \langle  v^{(2g-2)}_{(\beta,\gamma)}(\xi)  \prod_{\nu=1}^{\tilde n} \rho^1 (\gamma  ^ {-j_\nu - m _\nu  } ) (z_\nu ) \rho^1 (\bar  \gamma ^{- j_\nu - \bar m_\nu } )  (\bar z_\nu)  \right \rangle_{(\beta, \gamma)} \\
		& \quad \otimes 
		\left \langle  e^{ \frac{2 - 2 g}{b} \phi (\xi , \bar \xi) } \prod_{\nu=1}^{ n}  e^{2 b ( j_\nu  - \frac{1}{2b^2} )\phi (z_\nu , \bar z_\nu)}  \right \rangle  _{\phi}  \, .
	\end{aligned}
\end{align}
The $(\beta,\gamma)$-ghost part is computed as
\begin{align}
	&\left \langle v^{(2g-2)}_{(\beta,\gamma)}(\xi)   \prod_{\nu=1}^{\tilde n} \rho^1 (\gamma  ^ {-j_\nu - m _\nu  } ) (z_\nu ) \rho^1 (\bar  \gamma ^{- j_\nu - \bar m_\nu } )  (\bar z_\nu)  \right \rangle_{(\beta, \gamma)}= \delta^{(2)} \left(\sum_{\nu =1}^{ n}   m_\nu +\frac{k }{2} (2 R + n -2) \right) \nonumber  \\
	&\quad \times |\sigma (\xi)|^{4 g - 4} \prod_{\nu=1}^{\tilde n} ( E(\xi , z_\nu)^{2 - 2g} \sigma(z_\nu) ^{-2})^{ j_\nu + m_\nu +1 } ( E (\bar \xi , \bar  z_\nu)^{2 -2g} \sigma (\bar z_\nu)^{-2})^{j_\nu + \bar m_\nu +1}
	\label{onewindingg}
\end{align}
up to some normalization factors, see appendix \ref{sec:beta}.

The $\phi$-part of the correlator can be expressed in the path integral formulation as
\begin{align}
	\left \langle  e^{ \frac{2 - 2 g}{b} \phi (\xi , \bar \xi) }  \prod_{\nu=1}^{ n} e^{2 b( j_\nu - \frac{1}{2b^2}) \phi (z_\nu, \bar z_\nu)}  \right \rangle  _{\phi} 
	= \int \mathcal{D} \phi e^{-S[\phi]}  e^{ \frac{2 - 2 g}{b} \phi (\xi , \bar \xi) }  \prod_{\nu=1}^{ n} e^{2 b(  j_\nu - \frac{1}{2 b^2} )\phi (z_\nu , \bar z_\nu) }  \, , 
\end{align}
where the action is given by \eqref{Lactionsl2} with $Q_\phi = b$.
The field $\phi$ is then shifted by
\begin{align}
	\phi + \frac{1}{2b}  \left[ (2 - 2g)  \ln |E (\xi , z ) |^2  - 2 \ln |\sigma (z)|^2 + \ln |\rho (z)|^2  \right] \to \phi \, .
\end{align}
We further require the condition
\begin{align}
	(2 - 2g) \int_{z_0}^{\xi} \omega_l  + 2 \int_{(g-1)z_0}^\Delta \omega_l = 0  \label{xicond}
\end{align}
with $z_0$ being an arbitrary point in $\Sigma$, such that $\phi$ is periodic both before and after the shift.
With a similar analysis as in \cite{Hikida:2007tq,Hikida:2008pe}, we arrive at 
\begin{align}
	&\left \langle  e^{ \frac{2 - 2 g}{b} \phi (\xi , \bar \xi) }  \prod_{\nu=1}^{ n} e^{2 b( j_\nu - \frac{1}{2b^2}) \phi (z_\nu , \bar z_\nu) } \right \rangle  _{\phi}  \\
	&\quad = \int \mathcal{D} \phi e^{-S[\phi]}\prod_{\nu=1}^{ n} e^{2 b ( j_\nu - \frac{1}{2b^2}) \phi (z_\nu , \bar z_\nu) }  |E (\xi , z_\nu )^{2 g - 2} \sigma (z_\nu)^2|^{2 (j_\nu - \frac{1}{2b^2})} |\sigma (\xi)|^{4 - 4g} e^{\frac{3}{2} \left( \frac{1}{2b^2} -1 \right) U_g}\, . \nonumber 
\end{align}	
The action is \eqref{Lactionsl2} but with $Q_\phi = b - b^{-1}$ as in \eqref{Qnew}.
Overall, combining the two parts, we arrive at
\begin{align}
	\label{tildeA2g}
	\begin{aligned}
	\tilde A_n & =  \int \mathcal{D} \phi e^{-S[\phi]}\prod_{\nu=1}^n e^{2 b ( j_\nu - \frac{1}{2b^2})  \phi (z_\nu , \bar z_\nu) } \\
	& \quad \times \prod_{\nu=1}^{\tilde n}(E (\xi, z_\nu )^{2 g - 2 } \sigma (z_\nu)^{2})^{ - m_\nu - k/2} (E (\bar \xi , \bar z_\nu )^{2 g - 2} \sigma (\bar z_\nu)^{2})^{ - \bar m_\nu - k/2} 
	\end{aligned}
\end{align}
subject to the conditions \eqref{conserveg} and \eqref{conserve2}.
Here we have neglected the factor $e^{\frac{3}{2} \left( \frac{1}{2b^2} -1 \right) U_g}$.

The next step is again to consider a correlation function with a different insertion of vertex operators:
\begin{align}
	A_n \equiv \left \langle \prod_{\nu=1}^n V^{j_\nu, w_\nu}_{m_\nu , \bar m_\nu} ( z_\nu)   \prod_{\ell = 1}^ {R}V^{\frac{1}{2b^2},-1}_{\frac{k}{2},\frac{k}{2}}(y_\ell) \right \rangle  \label{Ang}
\end{align}
and relate it with \eqref{tildeAg} using the coset relation \eqref{spectralm}. 
On a Riemann surface of genus $g$, the condition for the sum of winding number is shifted as $|\sum_\nu w_\nu| \leq n -2 + 2g$, see, e.g., \cite{Hikida:2008pe}. In the presence of extra insertions at $z=y_\ell$, the condition is modified as
$|\sum_\nu w_\nu| -R \leq R + n -2 + 2g$, where the bound is saturated due to \eqref{Rg}.
%The conservation of $J_0^3$-charge is now
%\begin{align}
%	\sum_{\nu=1}^n m_\nu + \frac{k}{2} R = \sum_{\nu=1}^n \bar m_\nu + \frac{k}{2} R= - \frac{k}{2}( R + n -2) 
%\end{align}
%as in \eqref{conserve2}.  
As in the genus zero case, the correlation function \eqref{Ang} can be obtained as in \eqref{An}.
The $u(1)$ parts are given by \eqref{Bn} and 
\begin{align}
	\tilde B_n =\left \langle  e^{ (2g-2) \sqrt{\frac{k}{2}} \chi (\xi) }  \prod_{\nu=1}^n e^{ \sqrt{\frac{2}{k}} (m_\nu + \frac{k }{2} ) \chi (z_\nu)}  \prod_{\ell=1}^R e^{ \sqrt{2 k}  \chi (y_\ell) }  \right \rangle\, . 
\end{align}
A direct computation leads to
\begin{align}
	 B_n / \tilde B_n &= \Theta \prod_{\mu < \nu} E(z_\mu , z_\nu)^{ - m_\mu (w_\nu - 1) - m_\nu (w_\mu - 1) - \frac{k}{2 } (w_\mu w_\nu - 1 )   } 
	\prod_{\nu , \ell} E(z_\nu , y_\ell)^{2 (m_\nu + \frac{k }{2}) } \prod_{\ell < \ell '} E(y_{\ell} , y_{ \ell '})^{2 k} \nonumber \\
	& = \Theta  \prod_{\nu} (\tilde a_\nu ^\Gamma )^{ -  m_\nu - \frac{ k }{4}(\omega_\nu +1)} \prod_\ell \tilde \xi_\ell^{- \frac{k }{2} } 
\end{align}
with
\begin{align}
	\Theta = \prod_{\nu=1}^n E(\xi , z_\nu) ^{(2 g -2) (m_\nu + \frac{k}{2})} \prod_{\ell =1}^R E (\xi , y_\ell)^{ (2g -2)k} \, .
\end{align}

In this way, we arrive at
\begin{align}
	\begin{aligned}
		A_n & =  \prod_{\nu}  ( \tilde a_\nu ^\Gamma )^{ -  m_\nu -  \frac{k}{4} (w_\nu +1) }   ( \bar {\tilde a}{}_\nu ^\Gamma )^{ -  \bar m_\nu -  \frac{k}{4} (w_\nu +1) } \sigma (z_\nu)^{-2 m_\nu - k}\sigma (\bar z_\nu)^{-2 \bar m_\nu - k} \\ 
		& \quad \times \prod_\ell | \tilde \xi_\ell | ^{- k  } | \sigma (y_\ell) | ^{- 4 k  } \int \mathcal{D} \phi e^{-S[\phi]}\prod_{\nu=1}^n e^{2 b ( j_\nu - \frac{1}{2b^2})  \phi } (z_\nu) \, ,
	\end{aligned}
	\label{Anresultg}
\end{align}
where the action is given by \eqref{Lactionsl2} with $Q_\phi = b - b^{-1}$. 
As expected, the $\xi$-dependence disappears in the final expression.
As in the genus zero case, we conclude that  the correlation function \eqref{corrAdS} is given by \eqref{Anresultg} after removing the condition \eqref{conserve2}.
If we choose 
\begin{align}
	j_1 = \frac{1}{2b^2} (-1 + 2 g) + 1 -g \, , \quad j_\nu = \frac{1}{2b^2} \label{jg}
\end{align}
for $\nu = 2,3, \ldots ,n$,
then the path integral over $\phi$ yields only a constant factor.
The other factor in \eqref{Anresultg} reproduces \eqref{SymOrb0g} up to a constant by setting $m_\nu$ and $\bar{m}_\nu$ as in \eqref{hwinding}.
However, we should be careful  that $j_1$ in \eqref{jg} and $m_1$ in \eqref{hwinding} is not compatible with \eqref{onshell} for $g > 1$ in general. The exception is with $k=3$.
In other words,  we have shown the correspondence of correlation functions \eqref{corrSO} and \eqref{corrAdS} only for $k=3
$.  For the other cases, we need to consider more generic correspondence between \eqref{gcorrAdS} and \eqref{gcorrSO} as discussed at the end of section \ref{sec:sl2corr}.

\section{Conclusion and open problems} 
\label{sec:conclusion}

In this paper, we have studied the correspondence of correlation functions in the symmetric orbifold $\mathcal{M}^N/S_N$ and in the $sl(2)$ WZNW model. We first computed the correlation function of twist operators in the symmetric orbifold  from the partition function of $\mathcal{M}$ on the Riemann surface defined by the covering map $x = \Gamma(z)$. 
The correlation function can be summarized in a simple form as in \eqref{SymOrb0} and \eqref{SymOrb0g}.
We then examined the correlation function of dual operators  in the $sl(2)$ WZNW model.
For the case of sphere worldsheet, the correlation function \eqref{tildeA} was reduced to that of  Liouville field theory and the explicit form of \eqref{assumption3} was found by relating it to \eqref{tildeA} through a coset model.
We then showed that the essential part of \eqref{corrAdS} is obtained from \eqref{assumption3} and it reproduces the correlation function of twist operators in the symmetric orbifold when the conditions \eqref{jspecific} and \eqref{hwinding} are satisfied. We further generalize the analysis for the case of higher genus Riemann surface.

The correlation function \eqref{corrSO} in the symmetric orbifold corresponds to \eqref{corrAdS} in the $sl(2)$ WZNW model, where the vertex operators are given in the $x$-basis.
In this paper, we computed \eqref{assumption3}, where the vertex operators are in the $m$-basis. A relation between correlators \eqref{corrAdS} and \eqref{assumption3} was proposed. However, we could provide only indirect evidence for the relation by showing that both correlation functions satisfy the same equations with the help of the results in \cite{Eberhardt:2019ywk,Eberhardt:2020akk}.
As a consequence of it, we could examine the correspondence of correlation functions up to an overall factor. It is thus desired to obtain more direct ways to relate \eqref{corrAdS} and \eqref{assumption3}.

The correspondence of correlation functions analyzed in this paper should be a key ingredient to derive the AdS/CFT correspondence with superstrings on AdS$_3 \times S^3 \times T^4$ at the tensionless limit.
It seems to be straightforward to generalize the current analysis for the superstrings in the RNS formalism.
However, it was argued in \cite{Eberhardt:2018ouy} that the hybrid formalism of \cite{Berkovits:1999im} should be used to treat the limit in a proper way. This fact might make the analysis complicated, but we expect that the extension of reduction method developed in  \cite{Creutzig:2011qm} would be useful in this case.

It was argued in \cite{Gaberdiel:2013vva,Gaberdiel:2014cha} that a tensionless limit of AdS$_3$ superstrings is related to a higher spin AdS$_3$ gravity with $\mathcal{N}=4$ supersymmetry. 
It is an important task to reveal more direct relations including the symmetric orbifold CFT.
Moreover, it was also proposed that a matrix extension of AdS$_3$ higher spin gravity is dual to two dimensional CFT with $\mathcal{N}=3$ supersymmetry \cite{Creutzig:2011fe,Creutzig:2013tja,Creutzig:2014ula}. Matrix extended higher spin gravity is expected to describe stringy effects, see, e.g., \cite{Vasiliev:2018zer}.
It should be possible to derive correspondences of correlation functions associated with less supersymmetry as in \cite{Argurio:2000tg,Argurio:2000tb,Argurio:2000xm}.

\subsection*{Acknowledgements}

We are grateful to Pawel Caputa and Thomas Creutzig for useful discussions. 
The work of YH was supported by JSPS KAKENHI Grant Number 16H02182 and 19H01896.
The work of TL was supported by JSPS KAKENHI Grant Number 16H02182.

\appendix

\section{Correlation functions of ghost system}
\label{sec:beta}

In this appendix, we compute the correlation function
\begin{align}
	&\left \langle v^{(2g-2)}_{(\beta,\gamma)}(\xi)   \prod_{\nu=1}^{\tilde n} \rho^1 (\gamma  ^ {-j_\nu - m _\nu  } ) (z_\nu ) \rho^1 (\bar  \gamma ^{- j_\nu - \bar m_\nu } )  (\bar z_\nu)  \right \rangle_{(\beta, \gamma)} \, , \label{appcorr}
\end{align}
which appeared in \eqref{onewinding} and \eqref{onewindingg}.
The correlation function is in terms of  $\rho^1(\gamma^{-j-m})$, which are $\gamma$-ghosts with one unit of spectral flow. The correlator also includes the insertion of $v^{(2g-2)}_{(\beta,\gamma)}(\xi)$, which is obtained by acting $(2g-2)$ units of spectral flow to the identity operator.
Our strategy for the computation is to replace the roles of $\beta$ and $\gamma$ by making use of the automorphism $J^\pm \to - J^\mp$ and $J^3 \to - J^3$.  With this replacement, the correlation functions are given in terms of $\beta$ without the action of spectral flow. Moreover, the insertion of $v^{(2g-2)}_{(\beta,\gamma)}(\xi)$ can be treated by requiring a pole of order  $(2g-2)$ (or a zero of  second order for $g = 0$) at $z = \xi$ for $\beta(z)$ as in \cite{Hikida:2008pe}.

In order to find out the rule to replace $\beta$ and $\gamma$, we examine how the vertex operators change under $J^\pm \to - J^\mp$ and $J^3 \to - J^3$.  We are interested in vertex operators with one unit of spectral flow, which are characterized by the OPEs with the $sl(2)$ currents as
\begin{align}
	\begin{aligned}
		&J^+ (z) V^{j,1}_{m , \bar m} (0) = \frac{m +j}{z^2} V^{j,1}_{m +1 , \bar m} (0) + \mathcal{O} (z^{-1}) \, , \\
		&J^3 (z) V^{j,1}_{m , \bar m} (0) =\frac{m + k/2}{z} V^{j,1}_{m  , \bar m} (0) + \mathcal{O} (z^0) \, , \\
		&J^- (z) V^{j,1}_{m , \bar m} (0) = ( m  - j ) V^{j,1}_{m - 1 , \bar m} (0) + \mathcal{O} (z) \, .
	\end{aligned}
\end{align}
Now we perform the replacement of $J^\pm \to - J^\mp$ and $J^3 \to - J^3$.  
This also changes the quantum numbers of vertex operators as $ V^{j,1}_{m , \bar m} (0)  \to V^{j,-1}_{-m , -\bar m} (0)  $. 
Therefore, the OPEs between the $sl(2)$ currents and the vertex operators become
\begin{align}
	\begin{aligned}
	&	J^- (z) V^{j,-1}_{-m , -\bar m} (0) = \frac{- m - j}{z^2} V^{j,-1}_{-m -1 , \bar m} (0) + \mathcal{O} (z^{-1}) \, , \\
	&	J^3 (z) V^{j,-1}_{-m , -\bar m} (0) =\frac{- m - k/2}{z} V^{j,-1}_{-m  ,- \bar m} (0) + \mathcal{O} (z^0) \, , \\
	&	J^+ (z) V^{j,-1}_{-m ,- \bar m} (0) = ( - m  + j ) V^{j,-1}_{-m + 1 , -\bar m} (0) + \mathcal{O} (z) \, .
	\end{aligned}
\end{align}
These OPEs can be realized by
\begin{align}
	V^{j,-1}_{-m, -\bar m} (z) = N^j_{m,\bar m}\beta^{ j + m + 1} \bar \beta {}^{ j + \bar m + 1} e^{2 b (j - \frac{1}{2 b^2})\phi}
\end{align}
with a normalization factor $ N^j_{m,\bar m}$.
Therefore, the correlator \eqref{appcorr} can be identified with
\begin{align}
	\left \langle v^{(2g-2)}_{(\beta,\gamma)}(\xi)   \prod_{\nu=1}^{\tilde n} \beta (z_\nu) ^{j_\nu + m_\nu + 1 } \bar \beta (\bar z_\nu ) {}^{ j_\nu + \bar m_\nu  + 1 } \right \rangle_{(\beta , \gamma)}  \label{betacorr}
\end{align}
up to an overall factor.

The correlator \eqref{betacorr} is first evaluated for the $g=0$ case, followed by the extension to generic genus.
As mentioned above, $  v^{(-2)}_{(\beta,\gamma)}(\xi)  $ forces $\beta(z)$ to have a zero of second order at $z = \xi$.
Thus we can replace $\beta(z)$ by
\begin{align}
	\beta (z) = u (z - \xi)^2  \, . \label{betasol}
\end{align}
There is no condition to fix the overall factor $u$, therefore we decide to perform the integration over $u$.
Inserting \eqref{betasol} in \eqref{betacorr} with $g=0$, we have
\begin{align}
	&\left \langle  v^{(-2)}_{(\beta,\gamma)}(\xi)  \prod_{\nu=1}^{\tilde n} \beta(z_\nu )^{j_\nu + m_\nu +1 } \bar \beta (\bar z_\nu)^{j_\nu +\bar m_\nu + 1 } \right \rangle_{(\beta , \gamma)}   \label{betacorr2} \\
	&= \int \frac{d^2 u}{|u|^2} u^{\sum_{\nu=1}^{\tilde n} m_\nu + \frac{k}{2} (R + n -2)}
	\bar u {}^{\sum_{\nu=1}^{\tilde n} \bar m_\nu + \frac{k}{2} (R + n -2)} \prod_{\nu=1}^{\tilde n} (\xi - z_\nu)^{2 (j_\nu + m_\nu +1)} (\bar \xi - \bar  z_\nu)^{2 (j_\nu + \bar m_\nu +1)} \nonumber 
\end{align}
as in \eqref{onewinding}. The integration over $u$ gives a delta function, and  the measure of $u$ is chosen to be consistent with the ghost number violation.
% (... is chosen such that the ghost number criterion is not violated).

As in the case of $g=0$, we can obtain the expression of \eqref{betacorr} for $g > 0$ from the information of zeros and poles.
Due to the insertion of $ v^{(2g-2)}_{(\beta,\gamma)}(\xi) $, we set $\beta(z)$ to have a pole of order $(2g-2)$ at $z = \xi$ as
\begin{align}
	\beta (z) = \frac{u}{  E(z , \xi)^{2 g-2} \sigma (z)^{2} } \, .\label{betag}
\end{align}
Again we perform the integration over $u$ and fix its measure from the condition of ghost number violation. 
The dependence of $\sigma (z)$ is fixed such that  the solution of $\beta(z)$  subject to the condition \eqref{xicond} is periodic under the shifts along the homology cycles.

Inserting \eqref{betag} into \eqref{betacorr} with $g >0$, the correlation function has the correct type of poles and the dependence on $\sigma (z_\nu)$. However, the correlation function \eqref{betacorr} would also depend on $\sigma (\xi)$, and this dependence cannot be fixed in a similar way.
Therefore, we use a different formulation of $(\beta,\gamma)$-ghosts as
\begin{align}
	\beta (z) \simeq e^{- X(z)} \partial \xi (z) \, ,\quad \gamma (z) \simeq e^{X(z) }\eta (z) \, , 
\end{align}
where a bosonic field $X$ and fermionic fields $(\xi , \eta )$ satisfy
\begin{align}
	X(z) X(0) \sim - \ln z \, , \quad \xi (z)\eta (0) \sim \frac{1}{z} \, .
\end{align}
We also introduce their anti-holomorphic counterparts as $\bar X$ and $(\bar \xi  ,\bar \eta )$ and take a combination of the bosonic fields as 
$\mathcal{X} = X + \bar X$. Then the action for $\mathcal{X} $ is given by
\begin{align}
	S = \frac{1}{4 \pi} \int d^2 z \left(  \partial \mathcal{X} \bar \partial \mathcal{X} - \frac{1}{4} \sqrt{g} \mathcal{R} \mathcal{X} \right) \, . \label{xaction}
\end{align}
The twist operator $ v^{(2g-2)}_{(\beta,\gamma)}(\xi) $ has the correct OPEs with $\beta$ and $\gamma$ if we set
\begin{align}
 v^{(2g-2)}_{(\beta,\gamma)}(\xi)  \simeq e^{(2 - 2g)\mathcal{X} (\xi)} \, .
\end{align}
The insertion of $e^{(2 - 2g)\mathcal{X}  (\xi)}$ has a contraction with $\mathcal{X} $ in the second term of \eqref{xaction}, which yields a factor
\begin{align}
	\exp \left(  (2g -2) \frac{1}{16 \pi} \int d^2 z \sqrt{g (z) } \mathcal{R} (z) \ln |E ( \xi , z)|^2 \right) = |\sigma (\xi)|^{4g -4} \, .
\end{align}
In this way, we have obtained \eqref{onewindingg}.

%\bibliographystyle{JHEP}
%\bibliography{AdSCFT}

\providecommand{\href}[2]{#2}\begingroup\raggedright\endgroup

\end{document}